\documentclass[%
preprint,
showpacs,preprintnumbers,
nofootinbib,
bibnotes,
amsmath,amssymb,
aps
]{revtex4-1}

\usepackage{graphicx}
\usepackage{dcolumn}
\usepackage{bm}
\usepackage{amsfonts}
\usepackage{color}
\begin{document}


\title{Gravitational Wave Signatures from Low-mode Spiral Instabilities
\\ in Rapidly Rotating Supernova Cores}

\author{Takami Kuroda$^{1,2}$, Tomoya Takiwaki$^3$, and Kei Kotake$^{4,1}$}
\affiliation{$^1$Division of Theoretical Astronomy, National Astronomical Observatory of Japan, 
2-21-1, Osawa, Mitaka, Tokyo, 181-8588, Japan}
\affiliation{$^2$Department of Physics, University of Basel,
Klingelbergstrasse 82, 4056 Basel, Switzerland}
\affiliation{$^3$Center for Computational Astrophysics, National Astronomical Observatory of Japan, 2-21-1, Osawa, Mitaka, Tokyo, 181-8588, Japan}
\affiliation{$^4$Department of Applied Physics, Fukuoka University,
 8-19-1, Jonan, Nanakuma, Fukuoka, 814-0180, Japan}

\begin{abstract}
We study properties of gravitational waves (GWs) from rotating core-collapse of a 15$M_\odot$ star by performing three-dimensional
general-relativistic hydrodynamic simulations with an approximate neutrino transport. 
By parametrically changing the precollapse angular momentum, we focus on the effects of rotation on the
GW signatures in the early postbounce evolution.
Regarding three-flavor neutrino transport, we solve the energy-averaged set of radiation energy and 
momentum based on the Thorne's momentum formalism. In addition to
the gravitational quadrupole radiation from matter motions, we take into account GWs from anisotropic neutrino emission.
With these computations, our results present several supporting evidences
 for the previous anticipation that non-axisymmetric 
instabilities play an essential role in determining the postbounce 
GW signatures.
During prompt convection, we find that the waveforms show narrow-band and highly quasi-periodic 
signals which persist until the end of simulations. We point out that such feature reflects
the growth of the one-armed spiral modes.
 The typical frequency of the quasi-periodic waveforms can be well 
explained by the propagating acoustic waves between the stalled shock and the 
rotating proto-neutron star surface, which suggests the 
appearance of the standing-accretion-shock instability. Although the GW signals exhibit 
strong variability between the two polarizations and different observer directions, they are within the detection limits of next 
generation detectors such as by KAGRA and Advanced LIGO, if the source with sufficient angular momentum is located in our Galaxy.

\end{abstract}
\date{\today}

\maketitle

\section{Introduction}
\label{sec:Introduction}
The seminal paper \cite{EMuller82} on the gravitational-wave emission from stellar core-collapse started with the following two sentences:{\it``The current and 
near-future efforts (Weiss, 1979;Epstein, 1979;Weber, 1979; Douglass
and Braginsky, 1979)\cite{weiss,epstein,weber,douglass} to detect 
gravitational waves have increased the demand 
for reliable theoretical calculations of gravitational radiation 
from all possible energetic astrophysical sources. Among all 
potential sources, supernovae, possessing highly asymmetric cores and 
being situated in our own galaxy, have been thought to be 
the most promising candidates, probably already within the 
realm of second generation gravitational wave detectors.}"
It may be surprising that only by updating the references
in the late 1970's to the most recent ones (e.g., 
 Advanced LIGO/VIRGO 
\cite{Harry10,dega} and KAGRA \cite{somiya} for an observational side), 
above statements are true timelessly\footnote{although 
the most promising source for the first detection may be most likely for compact binary mergers..},
outlining the final goal toward which all the relevant studies regarding
the gravitational-wave (GW) 
signatures of core-collapse supernovae (CCSNe)
have been heading since then (see recent
reviews in \cite{Ott09,Kotake13}).

Following \cite{EMuller82}, most of the theoretical predictions have focused on the GW signals from rotational
collapse of the supernova cores (e.g., \cite
{Moenchmeyer91,yama95,Zwerger97,Kotake03,Kotake04,shibata04,Ott04,Ott07_prl,Ott07_cqg,Dimmelmeier02,Dimmelmeier07,Dimmelmeier08,Scheidegger08,Scheidegger10} and references therein).
In this context, rapid rotation of the precollapse core leads to
significant rotational flattening of the collapsing and bouncing core,
leading to a time-dependent quadrupole (or higher) GW emission.
 The steady improvement in the supernova models (e.g., the inclusion of a 
 micro-physical equation of state (EOS), general relativity (GR), treatment
 of neutrino physics) recently led to a theoretically best-studied, generic, so-called type I waveform of the bounce signals. 
The waveform is characterized by a 
 a sharp negative spike at bounce followed by a subsequent ring-down 
phase \citep{Dimmelmeier07,Ott07_prl,Ott07_cqg}.

After core-bounce, anisotropic matter motions due to convection 
\citep{Burrows96,EMuller97,Fryer04_apjl,EMuller04} and 
the Standing-Accretion-Shock-Instability (SASI, \citep{Marek09,Kotake07,Kotake09,kotake_ray,Murphy09}), and anisotropy in neutrino emission
\cite{EMuller04,Kotake09,EMuller12,BMuller13} 
are expected to be primary GW sources 
(see \cite{Ott09,Kotake13} for more details).
In general, these postbounce GW signals should be more 
difficult to detect compared to the bounce signals,
since they change much more stochastically with time as a result of 
chaotically growing convection and the SASI in the non-linear 
hydrodynamics (\cite{EMuller04,Kotake09,Marek09,Murphy09,Kotake11}). However an encouraging finding is recently 
reported by \cite{BMuller13} who made a detailed analysis on the 
GW signals from exploding two-dimensional (2D) models with
currently best available neutrino transport scheme. They pointed out 
that an episode of convective activities in the vicinity of the 
proto-neutron star (PNS) \cite{Murphy09} that trigger the onset of 
explosion is encoded in the GW 
spectrogram, which could be read out by a power-excess method 
\cite{Flanagan98}. 
A combined analysis of the emergent neutrino signals
\cite{Ott12a} should provide an important clue to the supernova engine
in the context of the neutrino-heating mechanism (e.g.,
\cite{Mezzacappa05,Janka12,Kotake12_ptep}), which is 
otherwise very hard to access for conventional astronomy by
electromagnetic waves (e.g., \cite{Kotake12}).

Regarding the bounce signals\footnote{We shortly call GW bursts 
emitted near bounce in the rotating iron cores as the 
bounce signals.}, rapid rotation is necessary for producing
distinct GW bursts and it is likely to obtain $\sim$ 1\% of massive star 
population (e.g., \cite{Woosley06}). Minor as they may be,
a high angular momentum of the precollapse core is essential 
 for the working of the magnetohydrodynamic 
(MHD) mechanism \cite{Bisnovatyi76,Leblanc70,Kotake06,Burrows07,Takiwaki09,
Takiwaki11}. 
This is because the MHD mechanism
relies on the extraction of rotational 
free energy of the collapsing core by means of 
the field-wrapping and magnetorotational instability (e.g., 
\cite{Balbus98,Obergaulinger09,Masada12} and references therein). 
It is worthwhile mentioning that such energetic MHD explosions 
are receiving great attention
 as a possible relevance to magnetars and collapsars
(e.g., \cite{Metzger11,MacFadyen01,Harikae09_a,Harikae10}), 
which are presumably linked to the formation of long-duration gamma-ray bursts (e.g., \cite{Meszaros06} for a review). The gravitational waveforms
from the MHD explosions were found to have
a quasi-monotonically increasing component that is produced by the 
bipolar outflows \cite{Shibata06,Obergaulinger06,Takiwaki11}.
 Such characteristic waveforms with
slower temporal evolution ($\lesssim 100$Hz)
are likely to be detectable by future space-based detectors like DECIGO 
\cite{Kawamura06}, which are free from seismic noises 
at the lower frequency bands.

Reliable modeling of rotating core collapse and 
the associated bounce signals should be done at least
in three-dimension (3D), because nonaxisymmetric instabilities are 
 expected to naturally develop in the postbounce supernova cores
(e.g., \cite{Andersson03,Fryer11} for reviews). 
Probably the best understood type of instability 
is a classical dynamical 
bar-mode instability with a threshold of $T/|W| \gtrsim 0.27$ 
\cite{Rampp98,Shibata05a}.
Here $T/|W|$ represents the ratio of rotational kinetic energy to gravitational energy. 
In contrast to the high $T/|W|$ instability, recent work,
some of which has been carried out in idealized setups and assumptions
\cite{Centrella01,Saijo03,Watts05,Ou06,Cerda-Duran07}
and later also in more self-consistent 3D simulations 
\cite{Ott05,Scheidegger08,Scheidegger10} 
including microphysical EOS and a prescribed (Liedend\"orfer's) 
deleptonization scheme
\cite{Liebendorfer05}, suggest 
that a differentially rotating PNS can become dynamically unstable
at much lower $T/|W|$ as low as $\lesssim 0.1$. 
Despite clear numerical evidence for their existence, the 
physical origin of the low-$T/|W|$ instability remains 
unclear. It is most probable \cite{Watts05,saijo06} that the 
instabilities are associated with the 
existence of corotation points (where the pattern speed of the unstable
modes matches the local angular velocity) inside the star and are thus 
likely to be a subclass of shear instabilities.\footnote{Note that corotation resonance has been long 
known to the key ingredients in the accretion disk system, such as the
Papaloizou-Pringle instability \cite{Papaloizou85}.}
 Even in the non-rotating case, it should be mentioned that 
the postbounce core is no longer
axisymmetric because of the growth of spiral SASI modes
\cite{Blondin07_nat,Iwakami08,Yamasaki08,Wongwathanarat10,Hanke13}, which is considered to be a 
key ingredient for explaining the origin of pulsar's spin \cite{Blondin07_nat}.

Since the vicinity of the PNS is an important cite for the 
non-axisymmetric instabilities, 
general relativity (GR) and accurate neutrino transport cannot be also 
neglected, not to mention 3D effects. 
In fact, Scheidegger et al.(2010)\cite{Scheidegger10} found in their 3D 
post-Newtonian models that 
the GW amplitudes become 5-10 times bigger for models including
deleptonization effects than those without. 
As is well known, the evolution and structure of the PNSs are very 
sensitive to the deleptonization episode via lepton/energy
transport from inside (e.g., \cite{Keil96,Bruenn96,Dessart06}). In the above case, the deleptonization 
 in the PNS results in the more 
compact and asymmetric core, leading to the stronger GW emission.
A deeper gravitational well in GR simulations \cite{BMuller13,Buras06a,KurodaT12} compared to the corresponding Newtonian models
should affect the criteria of the low-$T/|W|$ instability.

Despite the importance, recent multi-dimensional (multi-D)
models both including GR and relevant 
microphysics are mostly limited to 2D (e.g., \citep{Dimmelmeier08,BMuller08,BMuller13}),
and only handful 3D models have been so far reported \cite{Ott07_prl,Ott12a,Ott12b}.
Ott et al.~(2007) \cite{Ott07_prl} performed the first 
full 3D GR simulations employing a realistic EOS and 
Liedend\"orfer's deleptonization scheme \cite{Liebendorfer05}.
In one of their 3D models with the largest GW amplitudes \cite{Ott07_prl}, 
they reported the appearance of the low-$T/|W|$ instability 
after around $t_{\rm pb}\sim20$ ms postbounce,  
which is characterized by the dominant azimuthal density mode of $m=1$ \citep{Centrella01,Saijo03}. They found that the GW emission emitted in 
the direction of the spin axis become significantly stronger after 
the onset of the non-axisymmetric instability.
One of the limitations in \cite{Ott07_prl} is the use of
the deleptonization scheme that was originally 
designed to be valid only in the prebounce phase \cite{Liebendorfer05}. 
More recently, Ott et al. (2012) reported GR simulations employing a
neutrino leakage scheme with \cite{Ott12a} or without spatial symmetry
 assumptions in the computational domain \cite{Ott12b}, nevertheless, detailed 
analysis of the GW signals (including models with rapid rotation)
has not been reported yet.

In this work, we study rotating core-collapse and the bounce GW signals
 by performing fully general relativistic 3D simulations with 
an approximate neutrino transport.
The code is a marriage of an adaptive-mesh-refinement (AMR),
conservative 3D GR MHD code developed by 
\cite{KurodaT10}, and the approximate three-flavor neutrino transport code that was 
developed in our previous work (see \cite{KurodaT12} for details).
The spacetime treatment in the code
is based on the Baumgarte-Shapiro-Shibata-Nakamura (BSSN) 
formalism \citep[see, e.g.,][]{Shibata95,Baumgarte99}.
Regarding neutrino transport, we solve the energy-independent set of 
radiation moments
up to the first order and evaluate the second order momentum
with an analytic variable Eddington factor
(the so-called M1 closure scheme \citep{Levermore84}).
This part is based on the partial implementation of the Thorne's momentum formalism,  which was extended by \cite{Shibata11} in 
a more suitable manner applicable to the neutrino transport problem.
 By utilizing a nested grid 
infrastructure, an effective numerical 
resolution of our 3D models in the center 
is $\Delta x\sim 450$ m, which is as good as the most recent study 
by \cite{Ott12a} and is better than our 
previous work \cite{KurodaT12} ($\Delta x\sim 600$ m).
By parametrically changing the initial angular momentum 
 in the precollapse core of a 15 $M_{\odot}$ star \cite{WW95},
we follow the dynamics starting from the onset of gravitational 
collapse, through bounce, up to about 30-50 ms postbounce in this study.
 Albeit limited to the early postbounce phase (mainly due to the
computational cost), we will show several interesting GW features,
 which come from non-axisymmetric spiral waves that develop 
 under the influence of the advective-acoustic cycle
 which is characteristic to the (spiral) SASI.
 Our results indicate that the low-$T/|W|$ instability 
might play some roles in inducing non-axisymmetric spiral waves,
although we cannot unambiguously identify the existence of the low-$T/|W|$ instability limited 
by our shorter simulation time ($\lesssim 30$ ms postbounce) in this 
 work.

This paper is structured as follows. In Section
\ref{sec:Numerical Method},
we briefly summarize the numerical schemes, initial models, and the 
method how to extract
the gravitational waveforms. The main results are presented
in Section \ref{sec:Results}.
We summarize our results and discuss their implications in Section
\ref{sec:Summary and Discussions}.

\section{Numerical Method}
\label{sec:Numerical Method}
This section starts with a brief summary about the 
basic equations and the numerical schemes of GR radiation hydrodynamics.
We then move on to describe the initial conditions in Section 
\ref{initialmodels} and the method 
to extract the gravitational waveforms in Section 
\ref{Sec:Gravitational Wave Extraction}, respectively.

\subsection{GR Radiation Hydrodynamics}\label{first}
\subsubsection{Metric equations}\label{sec:metric}
The GR radiation hydrodynamic code developed in our previous 
work \cite{KurodaT12} consists of the following 
three parts, in which the evolution equations of metric, hydrodynamics, 
and neutrino radiation are solved, respectively (see, \cite{KurodaT12} for more details).
Each of them is solved in an operator-splitting manner, but the system evolves self-consistently
as a whole satisfying the Hamiltonian and momentum constraints.
Note in this section that geometrized units are used  
(i.e. both the speed of light and the gravitational constant are set to unity: $G = c = 1$).
Greek indices run from 0 to 3, Latin indices from 1 to 3.

Regarding the metric evolution, the spatial metric $\gamma_{ij}$ 
(in the standard (3+1) form: $ds^2=-\alpha^2dt^2+\gamma_{ij}(dx^i+\beta^idt)(dx^j+\beta^jdt),$ with $\alpha$ and $\beta^i$ being the lapse and shift,
respectively)
and its extrinsic curvature $K_{ij}$ are evolved using the BSSN 
formulation \citep{Shibata95,Baumgarte99}. The fundamental variables are
\begin{eqnarray}
 \phi &\equiv& \frac{1}{12}\ln[\det(\gamma_{ij})]\ , \\
 \tilde\gamma_{ij} &\equiv& e^{-4\phi}\gamma_{ij}\ , \\
 K &\equiv& \gamma^{ij}K_{ij}\ , \\
 \tilde A_{ij} &\equiv& e^{-4\phi}(K_{ij} - \frac{1}{3}\gamma_{ij}K)\ , \\
 \tilde\Gamma^i &\equiv& -\tilde\gamma^{ij}{}_{,j}\ .
\end{eqnarray}

The Einstein equation gives rise to the evolution equations for the BSSN variables as,
\begin{eqnarray}
\label{eq:BSSN1}
(\partial_t-\mathcal{L}_\beta)\tilde\gamma_{ij}&=&-2\alpha\tilde A_{ij} \\
\label{eq:BSSN2}
(\partial_t-\mathcal{L}_\beta)\phi&=&- \frac{1}{6}\alpha K \\
\label{eq:BSSN3}
(\partial_t-\mathcal{L}_\beta)\tilde A_{ij}&=&e^{-4\phi}\left[ \alpha (R_{ij} -8\pi \gamma_{i\mu}\gamma_{j\nu}T^{\mu\nu}_{\rm (total)})\right.\nonumber \\
&&\left. -D_iD_j \alpha\right]^{\rm trf}\nonumber+\alpha(K\tilde A_{ij}-2\tilde A_{ik}\tilde \gamma ^{kl}\tilde A_{jl}) \nonumber \\ \\
\label{eq:BSSN4}
(\partial_t-\mathcal{L}_\beta)K&=&-\Delta \alpha +\alpha (\tilde A_{ij}\tilde A^{ij}+K^2/3)\nonumber \\
&&+4\pi \alpha (n_\mu n_\nu T^{\mu\nu}_{\rm (total)}+\gamma^{ij}\gamma_{i\mu}\gamma_{j\nu}
T^{\mu\nu}_{\rm (total)}) \nonumber \\ \\
\label{eq:BSSN5}
(\partial_t-\beta^k\partial_k)\tilde\Gamma^i&=&16\pi\tilde\gamma^{ij}\gamma_{i\mu}n_\nu 
T^{\mu\nu}_{\rm (total)} \nonumber\\
&&-2\alpha(\frac{2}{3}\tilde\gamma^{ij}K_{,j}-6\tilde A^{ij}\phi_{,j}-\tilde\Gamma^i_{jk}\tilde A^{jk})\nonumber\\
&&+\tilde\gamma^{jk}\beta^i_{,jk}+\frac{1}{3}\tilde\gamma^{ij}\beta^k_{,kj}-\tilde\Gamma^{j}\beta^i_{,j}\nonumber\\
&&+\frac{2}{3}\tilde\Gamma^{i}\beta^j_{,j}+\beta^j\tilde\Gamma^i_{,j}-2\tilde A^{ij}\alpha_{,j},
\end{eqnarray}
where ${\mathcal{L}}_{\beta}$ is the Lie derivative with respect to $\beta^i$,
$T^{\mu \nu}_{\rm (total)}$ is the total stress-energy tensor 
(fluid + radiation which we shall discuss in the next subsection),
$D$ denotes covariant derivative operator associated 
with $\gamma_{ij}$, $\Delta=D^iD_i$,``trf'' denotes the trace-free operator, 
$n_\mu=(-\alpha,0)$ is the time-like unit vector normal to 
the $t=$ constant time slices.
Following  \cite{Alcubierre01}, the gauge is specified by the 1+log lapse, 
\begin{equation}
\label{eq:1+log}
\partial_t\alpha=\beta^i\partial_i\alpha-2\alpha K,
\end{equation}
and by the Gamma-driver-shift,
\begin{equation}
\label{eq:GammaDriver}
\partial_t\beta^i=k\partial_t \tilde\Gamma^i,
\end{equation}
here we chose $k=1$.

\subsubsection{Radiation Hydrodynamics}
\label{sec:Radiation Hydrodynamics}
The total stress-energy tensor $T^{\alpha\beta}_{\rm (total)}$ appeared in 
Equations (8-10) is expressed as
\begin{equation}
T_{\rm (total)}^{\alpha\beta} = T_{\rm (fluid)}^{\alpha\beta} + \sum_{\nu\in\nu_e,\bar\nu_e,\nu_x}T_{(\nu)}^{\alpha\beta}
\label{TotalSETensor}
\end{equation}
where $T_{\rm (fluid)}^{\alpha\beta}$ and $T_{(\nu)}^{\alpha\beta}$ is the 
stress-energy tensor of fluid and neutrino radiation field, respectively.
Note in the above equation, summation is taken for 
all species of neutrinos ($\nu_e,\bar\nu_e,\nu_x$) with $\nu_x$ 
representing heavy-lepton neutrinos (i.e. 
$\nu_{\mu}, \nu_{\tau}$ and their anti-particles).

Starting from the definition of $T_{(\nu)}^{\alpha\beta}$,
\begin{eqnarray}
{T_{(\nu)}}^{\alpha\beta} \equiv E_{(\nu)} n^\alpha n^\beta+ F_{(\nu)}^\alpha n^\beta+
F_{(\nu)}^\beta n^\alpha +
P_{(\nu)}^{\alpha\beta},
\label{t_lab}
\end{eqnarray}
the evolution equations of radiation energy ($E_{(\nu)}$) and 
radiation flux ($F^{\alpha}_{(\nu)}$) in 
 the truncated momentum 
formalism \cite{Thorne81} can be expressed as \cite{Shibata11},
\begin{eqnarray}
\partial_t (e^{6\phi}E_{(\nu)})+\partial_i [e^{6\phi}(\alpha F_{(\nu)}^i-\beta^i E_{(\nu)})] =\nonumber \\
e^{6\phi}(\alpha P^{ij}K_{ij}-F_{(\nu)}^i\partial_i \alpha-\alpha Q^\mu n_\mu),
\label{rad1}
\end{eqnarray}
and 
\begin{eqnarray}
\partial_t (e^{6\phi}{F_{(\nu)}}_i)+\partial_j [e^{6\phi}(\alpha 
{P_{(\nu)}}_i^j-\beta^j {F_{(\nu)}}_i)] =\nonumber \\
e^{6\phi}[-E_{(\nu)}\partial_i\alpha +{F_{(\nu)}}_j\partial_i \beta^j+(\alpha/2) 
P_{(\nu)}^{jk}\partial_i \gamma_{jk}+\alpha Q^\mu \gamma_{i\mu}],\nonumber \\
\label{rad2}
\end{eqnarray}
where $Q^{\mu}$ denotes the source terms. In order to simplify the 
neutrino-matter interactions, a methodology of neutrino leakage scheme 
is partly employed at this moment 
(see \citep{KurodaT12} for the explicit expressions).

By adopting the M1 closure \citep{Levermore84}, the radiation pressure is expressed as
\begin{eqnarray}
{P_{(\nu)}}^{ij}=\frac{3\chi-1}{2}P^{ij}_{\rm thin}+\frac{3(1-\chi)}{2}P^{ij}_{\rm 
thick},
\label{neu_p}
\end{eqnarray}
where $\chi$ represents the variable Eddington factor,
$P^{ij}_{\rm thin}$ and $P^{ij}_{\rm thick}$ correspond to the radiation pressure in the optically thin and
thick limit, respectively.
They are written in terms of $E_{\nu}$ and $F_{\nu,i}$ \citep{Shibata11}.
For the variable Eddington factor $\chi$, we employ the one proposed by 
\cite{Levermore84},
\begin{eqnarray}
\label{eq:Livermore_chi}
\chi&=&\frac{3+4\bar{F}^2}{5+2\sqrt{4-3\bar{F}^2}},\\
\bar{F}^2&\equiv&\frac{F^iF_i}{E^2}.
\end{eqnarray}

The hydrodynamic equations are written in a conservative form as,
\begin{eqnarray}
\label{eq:GRmass}
\partial_t \rho_{\ast}&+&\partial_i(\rho_\ast v^i)=0,\\
\label{eq:GRmomentum}
\partial_t \hat S_i&+&\partial_j(\hat S_i v^j+\alpha e^{6\phi}P\delta_i^j)=\nonumber \\
&&-\hat S_0\partial_i \alpha+\hat S_k\partial_i \beta^k+2\alpha e^{6\phi}S_k^k\partial_i \phi\nonumber \\
&&-\alpha e^{2\phi} ({S}_{jk}-P \gamma_{jk}) \partial_i 
\tilde{\gamma}^{jk}/2-e^{6\phi}\alpha Q^\mu \gamma_{i\mu},\nonumber\\ \\
\label{eq:GRenergy}
\partial_t \hat \tau&+&\partial_i (\hat S_0v^i+e^{6\phi}P(v^i+\beta^i)-\rho_\ast v^i)=\\
&&\alpha e^{6\phi} K S_k^k /3+\alpha e^{2\phi} ({S}_{ij}-P \gamma_{ij})\tilde{A^{ij}}\nonumber\\
&&-\hat S_iD^i\alpha+e^{6\phi}\alpha Q^\mu n_\mu,\\
\label{eq:GRlepton}
\partial_t (\rho_\ast Y_e)&+&\partial_i (\rho_\ast Y_e v^i)=\rho_\ast \Gamma_e,
\end{eqnarray}
where $\hat X\equiv e^{6\phi}X$, $\rho_\ast\equiv\rho We^{6\phi}$, $S_i\equiv\rho hW u_i $ and
$S_0\equiv\rho h W^2-P$.
$\rho$ is the rest mass density, $u_\mu$ is the 4-velocity of fluid, $h\equiv 1+\varepsilon+P/\rho$ is the specific enthalpy,
$v^i=u^i/u^t$, $\hat\tau=\hat S_0-\rho_{\ast}$, $Y_e$ is the electron fraction,
$\varepsilon$ and $P$ is the internal energy and pressure, respectively.
We employ the Shen EOS \cite{Shen98} for baryonic matter which is based on the relativistic mean field theory (see \citep{KurodaT12} for more details on the EOS implementation).

\subsection{Initial Models}
\label{initialmodels}
We employ a widely used progenitor of a 15$M_\odot$ star 
(\cite{WW95}, model ``s15s7b2'') and impose precollapse rotation
manually to study its effect during the collapse, bounce, and 
the early postbounce phases in a controlled fashion.
We assume a cylindrical rotation profile,
\begin{eqnarray}
\label{eq:rotlaw}
u^t u_\phi ={\varpi_0}^2(\Omega_0-\Omega),
\end{eqnarray}
where $u_\phi = \sqrt{u_x^2+u_y^2}$ is rotational 
velocity and $\varpi_0$ set here to be $10^8$cm is reconciled
with results from stellar evolution calculations suggesting uniform 
rotation in the precollapse core. The initial angular velocity at the origin $\Omega_0$ is 
treated as a free 
parameter and we compute four models by varying 
$\Omega_0=0,\ \pi/6,\ \pi/2$ and $\pi$ (rad s$^{-1}$). We enumerate 
models as R$n$, where $n$ ranges from 0 (non-rotating) to 3 (rapidly
rotating), that corresponds to the four choices of the precollapse
central velocity (see Table \ref{tab:CoreBounce}).
Bearing in mind that it is very hard to go beyond 1D computations
in stellar evolution calculations and 
their outcomes may not be the final answer, the central angular velocity
for a 15$M_\odot$ star is predicted to range from 0.15 \cite{Heger05} 
to 3 (rad/s) \cite{Heger00}, depending on
whether the (prescribed) angular momentum transport is included or not \cite{Spruit98}.
 The central angular velocity for models R1 and R3 is adjusted to be 
closely in the same range  with the progenitor models (e.g.,
Table \ref{tab:CoreBounce}).
We add a random 1\% initial density 
perturbation for all the models. By doing so, we hope to see 
convective activities shortly after bounce in our 3D-GR simulations 
including neutrino transport (albeit approximate) that are currently 
able to follow at most $\sim $50 ms postbounce for each model (limited by the currently available supercomputing power at our hand).
 For model R3, we also computed two more models with
different conditions as, without seed perturbation R3$_{\rm woP}$ and
with switching off the neutrino-matter interaction after the bounce R3$_{\rm off}$ and {\bf one more model without seed perturbation
 (model $R3_{\rm delep}$), in which a deleptonization scheme \cite{Liebendorfer05} is employed exactly as in \cite{Ott07_prl}}.

The outer boundary is at the radius of 7500 km and 
nested boxes with 9 refinement level are embedded
in the computational domain without any spatial symmetry.
Each box contains $128^3$ cells and the minimum grid size at the origin is thus $\Delta x=450$m
which is the same resolution with \cite{Ott12a} and better than our previous study with $\Delta x=600$m \citep{KurodaT12}.
In the vicinity of standing shock front $R\sim120(240)$ km, our resolution achieves $\Delta x=1.8(3.6)$ km
which is a factor of 2(4) coarser than \cite{Ott12a}.

\subsection{Gravitational Wave Extraction}
\label{Sec:Gravitational Wave Extraction}
From GR simulations, GWs are extractable either 
by gauge-dependent or -independent methods.
\cite{Shibata05a} compared the waveforms emitted from oscillating
neutron stars by the two methods and reported the quadrupole formula produces the waveform in a sufficient accuracy
compared to the gauge-invariant method.
They showed the quadrupole formula underestimates the wave amplitude by $\sim10$\%, but the phase is not changed drastically. More detailed comparison is recently reported in \cite{Reisswig11}, 
which also supports that the quadrupole approximation is adequate 
for stellar collapse spacetimes with a PNS formation.
Accordingly in this work we choose the conventional quadrupole formula 
\cite{Scheidegger10,Sekiguchi10,Shibata05a,Kotake04} for extracting 
GWs from our simulations.

In the quadrupole formula, the transverse and the trace-free gravitational field $h_{ij}$ is expressed by \cite{Misner73,Finn90}
\begin{eqnarray}
\label{eq:hij}
h_{ij}=\frac{A_+e_++A_\times e_\times}{D}
\end{eqnarray}
In Eq.(\ref{eq:hij}), $A_{+/\times}$ represent amplitude of orthogonally polarized wave components, $e_{+/\times}$ denote unit polarization tensors and $D$ is the source distance.
Following \cite{Scheidegger08}, we adopt the same expressions for the
wave amplitude $A_{+/\times}$ as,
\begin{eqnarray}
\label{eq:A+}
A_+(\theta,\phi)&=&\ddot I_{\theta \theta}^{TT}-\ddot I_{\phi \phi}^{TT} \\
\label{eq:Ax}
A_\times(\theta,\phi)&=&2\ddot I_{\theta \phi}^{TT},
\end{eqnarray}
where the quadrupole moment in the spherical coordinates $\ddot
I_{\theta \theta}$, $\ddot I_{\phi \phi}$, and $\ddot I_{\theta \phi}$
are expressed into the Cartesian components $\ddot I_{ij}$ as \cite{Oohara97,Scheidegger08}
\begin{eqnarray}
\label{eq:Ithetatheta}
\ddot I_{\theta \theta}^{TT}&=&(\ddot I_{xx}^{TT}\cos^2\phi+\ddot I_{yy}^{TT}\sin^2\phi+2\ddot I_{xy}^{TT}\sin \phi \cos\phi)\cos^2\theta\nonumber \\
&&+\ddot I_{zz}^{TT}\sin^2\theta -2(\ddot I_{xz}^{TT}\cos\phi+\ddot I_{yz}^{TT}\sin\phi)\sin\theta\cos\theta,\nonumber\\ \\
\label{eq:Iphiphi}
\ddot I_{\phi \phi}^{TT}&=&\ddot I_{xx}^{TT}\sin^2\phi+\ddot I_{yy}^{TT}\cos^2\phi-\ddot I_{xy}^{TT}\sin 2\phi, \\
\label{eq:Ithetaphi}
\ddot I_{\theta \phi}^{TT}&=&(\ddot I_{yy}^{TT}-\ddot I_{xx}^{TT})\cos\theta\sin\phi\cos\phi\nonumber \\
&&+\ddot I_{xy}^{TT}\cos\theta(\cos^2 \phi-\sin^2\phi)\nonumber \\
&&+\ddot I_{xz}^{TT}\sin\theta\sin\phi -\ddot I_{yz}^{TT}\sin\theta\cos\phi,
\end{eqnarray}
with superscripts ${TT}$ denoting projection into the transverse-traceless gauge.
The quadrupole moment $I_{ij}$ and its time derivative are given as \cite{Shibata05a},
\begin{eqnarray}
\label{eq:Iij}
I_{ij}=\frac{G}{c^4}\int \rho_\ast x^i x^j d^3x,
\end{eqnarray}
and
\begin{eqnarray}
\label{eq:dIijdt}
\dot I_{ij}=\frac{G}{c^4}\int \rho_\ast (v^i x^j+x^i v^j) d^3x.
\end{eqnarray}
The second time derivative $\ddot I_{ij}$ is directly computed by
 taking the time derivative of Eq.(\ref{eq:dIijdt}) numerically.

According to \cite{EMuller97,Kotake11},  we estimate GWs from
anisotropic neutrino radiation $A^\nu_{+/\times}$ as
\begin{eqnarray}
\label{eq:A+Nu}
A^\nu_+(\xi,t)&=&\frac{2G}{c^4}\int_0^t dt' \int d\Omega' (1+s(\theta')c(\phi')s(\xi)+c(\theta')c(\xi))\nonumber \\
&&\times \frac{[s(\theta')c(\phi')c(\xi)-c(\theta')s(\xi)]^2-[s(\theta')s(\phi')]^2}
{[s(\theta')c(\phi')c(\xi)-c(\theta')s(\xi)]^2+[s(\theta')s(\phi')]^2}\nonumber \\
&&\times R_{L_\nu}^2 {F^r_{(\nu)}}'
\end{eqnarray}
and
\begin{eqnarray}
\label{eq:AxNu}
A^\nu_\times(\xi,t)&=&\frac{2G}{c^4}\int_0^t dt' \int d\Omega' (1+s(\theta')c(\phi')s(\xi)+c(\theta')c(\xi))\nonumber \\
&&\times \frac{s(\theta')s(\phi')[s(\theta')c(\phi')c(\xi)-c(\theta')s(\xi)]}
{[s(\theta')c(\phi')c(\xi)-c(\theta')s(\xi)]^2+[s(\theta')s(\phi')]^2}\nonumber \\
&&\times R_{L_\nu}^2 {F^r_{(\nu)}}'.
\end{eqnarray}
Here $s(A)\equiv {\rm sin}(A)$, $c(A)\equiv {\rm cos}(A)$ and ${F^r_{(\nu)}}'$ is radial energy flux of each neutrino species
($\nu = \nu_e,\bar\nu_e,\nu_x$) estimated at the radius of 
$(R_{L_\nu},\theta',\phi')$.
$\xi$ denotes observer angle relative to the rotational axis and $\xi=0$ and $\pi/2$ for polar and equatorial observers, respectively.

\section{Results}
\label{sec:Results}
\subsection{Hydrodynamic features}
\label{sec:Hydrodynamic Evolution}
\begin{figure}[hbpt]
\begin{center}
\includegraphics[width=170mm,angle=-90]{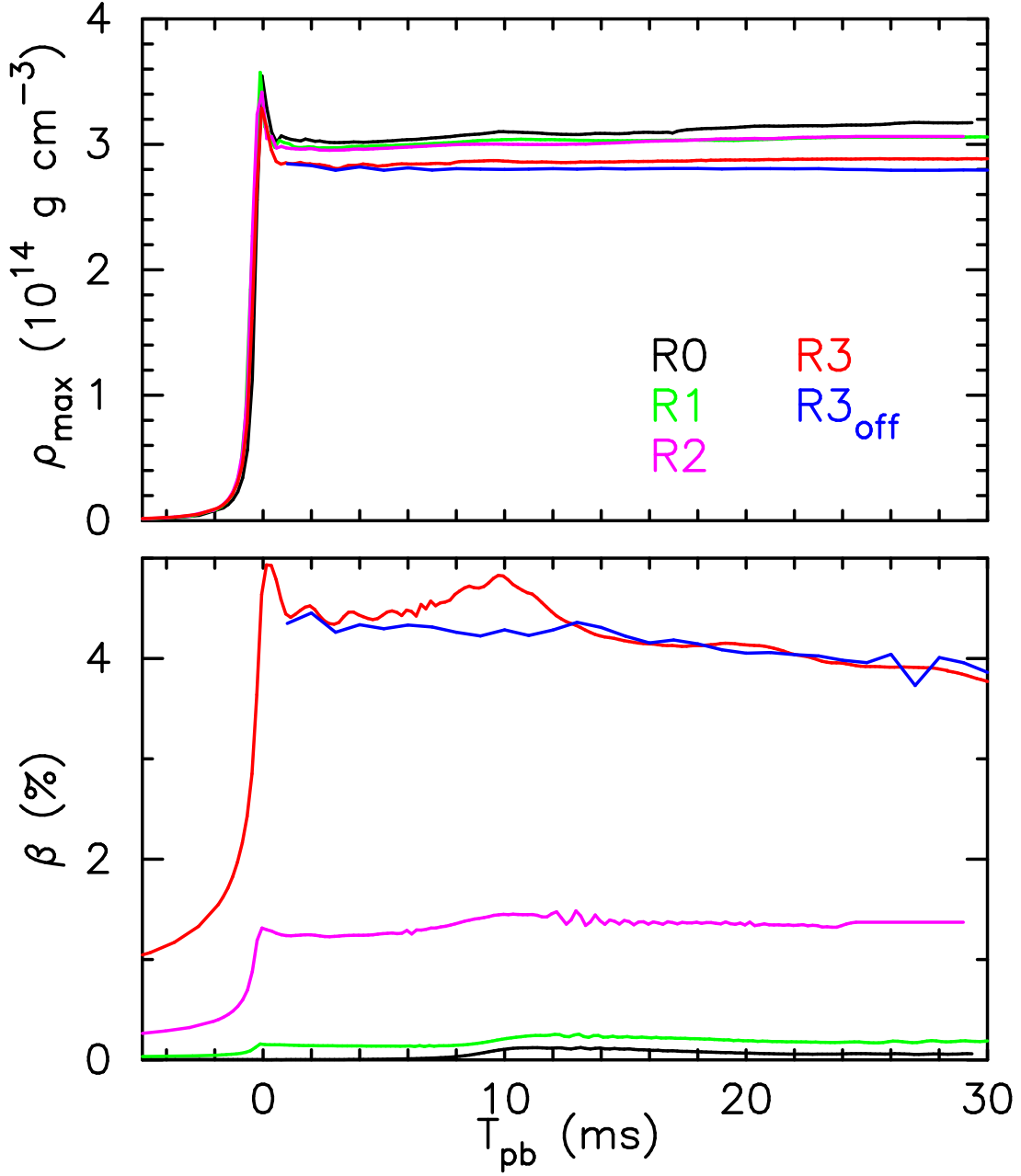}
\end{center}
\caption{\label{pic:diag} Time evolution of the
maximum rest mass density ({\it top}) and the rotational $\beta$
 parameter ({\it bottom}) as a function of postbounce
 time ($T_{\rm pb}$) for models R0(black line), R1(green line),
 R2(magenta line), R3(red line) and R3$_{\rm off}$(blue line), respectively.}
\end{figure}

To summarize hydrodynamic features of our models, we first show
 evolution of the central (rest-mass) density and
rotational $\beta$ parameter (i.e., the ratio of rotational kinetic
energy ($T$) to gravitational potential energy ($W$)
near bounce in Fig.\ref{pic:diag}
(see also table \ref{tab:CoreBounce} in which several key quantities are summarized).
Note that $T$ and $W$ is respectively defined as \cite{Kiuchi08}
\begin{eqnarray}
T\equiv\frac{1}{2}\int \rho_\ast h u_\phi v^\phi dx^3,
\end{eqnarray}
where $v^i=u^i/u^t$ and
\begin{eqnarray}
\label{eq:gravitational-potential}
W\equiv M_{\rm bar}-M_{\rm ADM}+E_{\rm int}+E_{\rm kin}+E_{\rm rad},
\end{eqnarray}
 with $M_{\rm bar}$, $M_{\rm ADM}$, $E_{\rm int}$, $E_{\rm kin}$ and $E_{\rm rad}$
 being the total baryon mass, ADM mass, internal energy, kinetic energy
 and neutrino radiation energy, respectively\footnote{
 Following an analogy of electro-magnetic energy \cite{Kiuchi08},
 we evaluated the contribution from neutrino radiation field, $E_{\rm rad}$, as
\begin{eqnarray}
\label{eq:neutrino radiation energy}
E_{\rm rad}&\equiv& \int T_{(\nu)}^{\alpha\beta}n_\alpha u_\beta \sqrt{\gamma} dx^3\nonumber \\
&=&\int (E_{(\nu)} W- F_{(\nu)}^i u_i) \sqrt{\gamma} dx^3.
\end{eqnarray}}.
 The top panel of Fig.1 shows that the central densities for all
  the models at bounce exceed nuclear density ($\sim 2.8\times 10^{14}~{\rm g}{\rm cm}^{-3}$). As consistent with
  \cite{Ott07_prl,Ott12a}, multiple bounce does not appear also in this study,
   which was otherwise often observed in previous
   Newtonian simulations with simplified setups.
   From the bottom panel of Fig.1, the $\beta$ parameter at
 bounce reach 0.15, 1.3 and 4.9 \% for models R1, R2, and R3,
 which is consistent with \cite{Ott12a}
 (hereafter Ott+12). Ott+12 reported full 3D GR simulations with
 relatively similar
 schemes as ours, i.e. the BSSN formalism including
neutrino leakage (but without neutrino heating). The precollapse 
 density structure of a 12 $M_{\odot}$ star \cite{Woosley07}
employed in \cite{Ott12a} is similar to the one employed 
 in this work (i.e., the 15 $M_{\odot}$ star \cite{WW95}). Since
 our model series of R0 - R3 employ similar initial central angular
 velocity as their ``s12WH07j(0-3)" in order, 
it is suitable to make comparisons.
For instance, their $\beta$ parameters for ``s12WH07j(1-3)"
 at bounce are 0.4, 1.6, and 5.1 \%, which are in good agreement with
   our counterpart models (R1-R3).
On the other hand, there exist some differences in numerical setups between
 our study and Ott+12, including the adopted EOS, 
initial perturbations, and spatial restrictions.  For example,
the increase in the central density over the first 30 ms postbounce 
(e.g., Figure 3 in \cite{Ott12a}) is slightly milder for our models. 
This is because the Shen EOS employed in this work is stiffer than the 
Lattimer-Swesty EOS (the nuclear incompressibility $K=220$ MeV) 
employed in \cite{Ott12a}.
Several important
 impacts on the GW emission will be mentioned later.

 \begin{table}[b]
\caption{\label{tab:CoreBounce}%
Model summary. The second column represents the initial
 central angular velocity ($\Omega_{\rm ini}$) followed by
$\rho_{\rm max,b}$ (the third column) and $\beta_{\rm b}$ (the fourth
   column), which represents the maximum rest mass
 density and rotational $\beta$ parameter at bounce, respectively.
}
\begin{ruledtabular}
\begin{tabular}{cccc}
\textrm{Model}&
\textrm{$\Omega_{\rm ini}$ (rad s$^{-1}$)}&
\textrm{$\rho_{\rm max,b}(10^{14} {\rm g\ cm}^{-3})$}&
\textrm{$\beta_{\rm b}$}\\
\colrule
R0& 0 & 3.54 & $2.3\times10^{-5}$\\
R1&$\pi/6$ & 3.52 & $1.5\times10^{-3}$\\
R2&$\pi/2$ & 3.41 & $1.3\times10^{-2}$ \\
R3 & $\pi$ & 3.28 & $4.9\times10^{-2}$\\
\end{tabular}
\end{ruledtabular}
 \end{table}

\begin{figure*}[hbpt]
\begin{center}
\includegraphics[width=75mm]{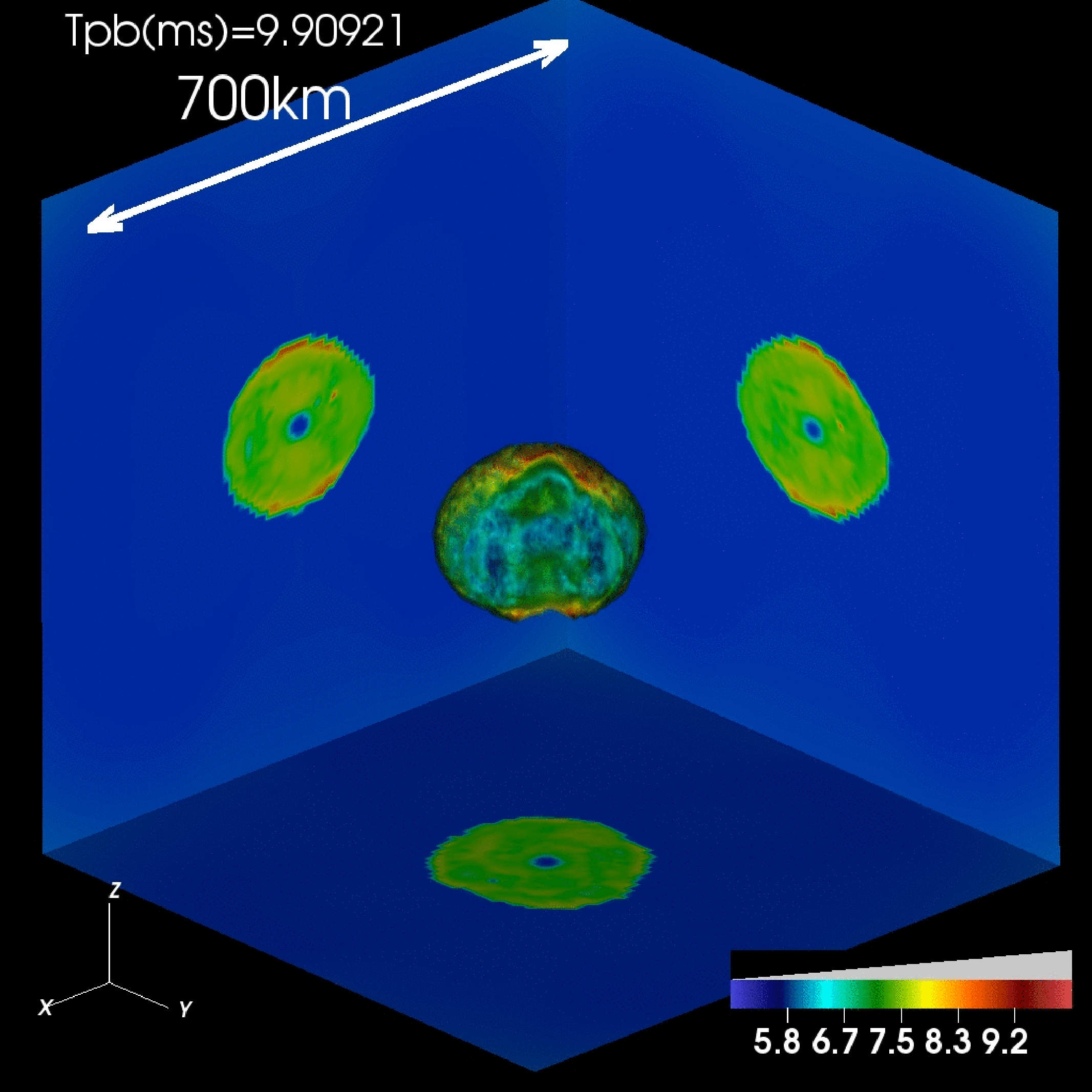}
\includegraphics[width=75mm]{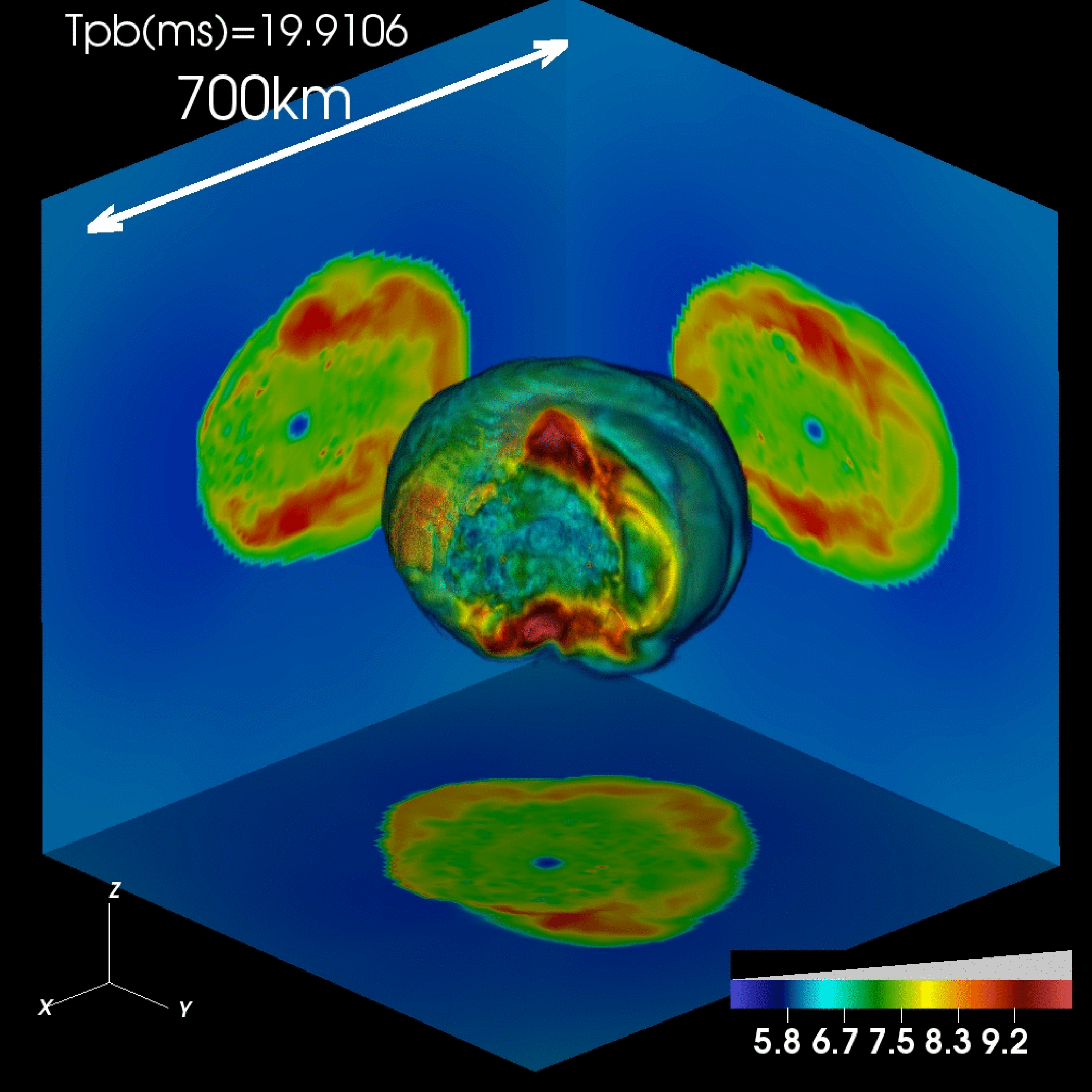}\\
\includegraphics[width=75mm]{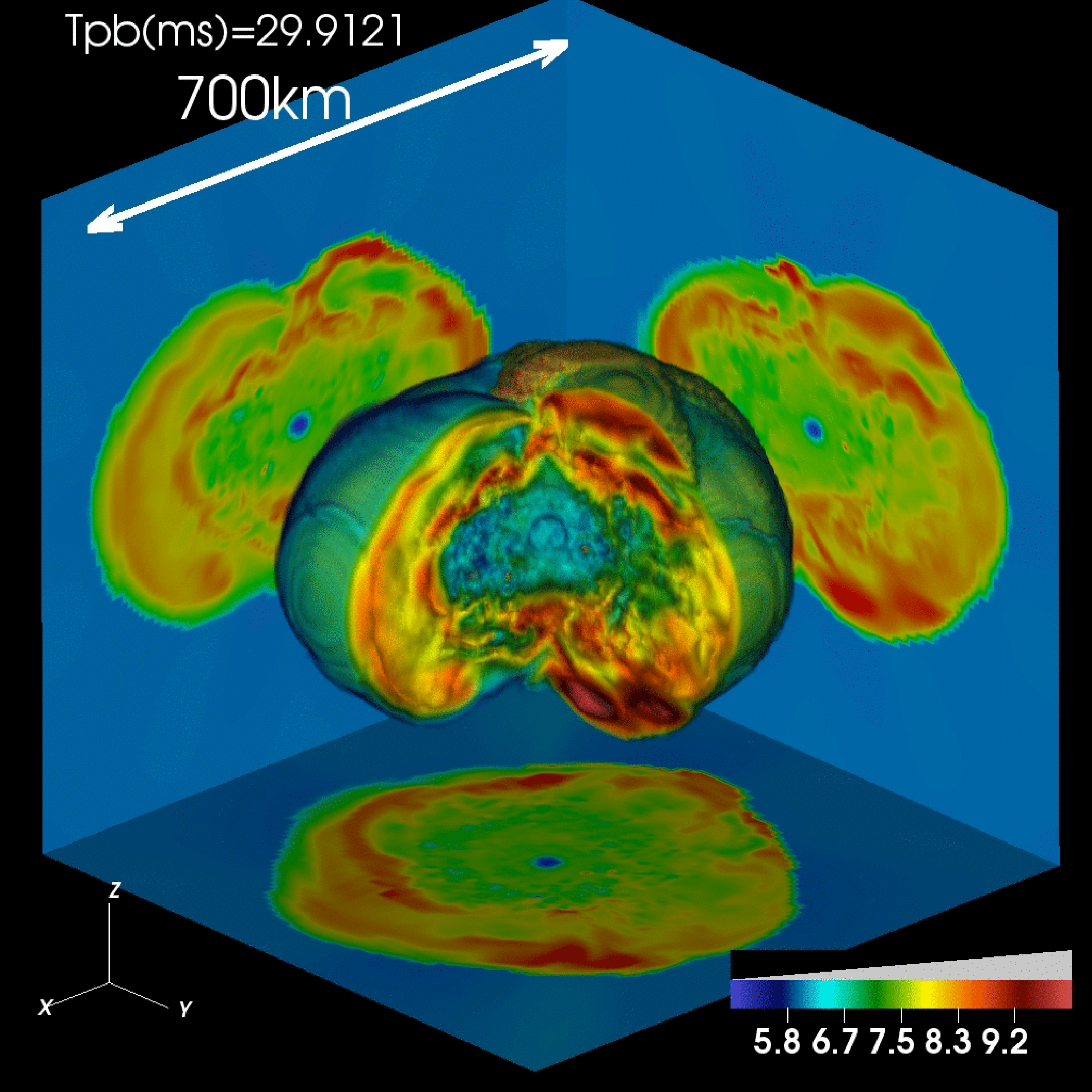}
\includegraphics[width=75mm]{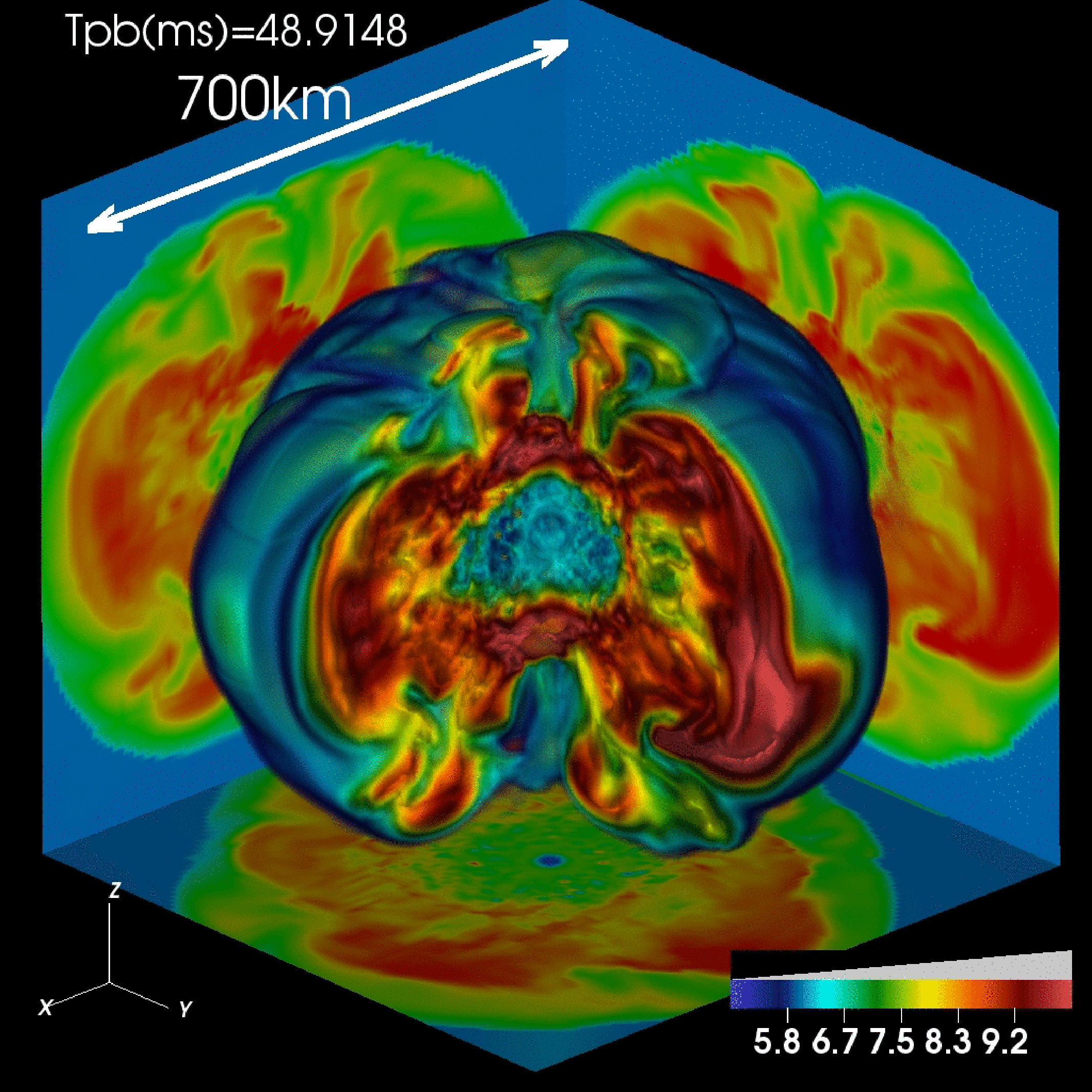}
\end{center}
\caption{\label{pic:3Dfigure_1} Several snapshots of entropy distributions
 ($k_{\rm B}$ baryon$^{-1}$)
in the central cube of 700km$^3$ for model R3
(top left; $t_{\rm pb}=9.9$ ms, top right; $t_{\rm pb}=19.9$ ms, bottom left; $t_{\rm pb}=29.9$ ms and bottom right; $t_{\rm pb}=48.9$ ms).
The contours on the cross sections in the $x$ = 0 (back right), $y$ = 0 (back left), and $z$ = 0 (bottom) planes are, respectively projected on the sidewalls of the graphs to visualize 3D structures.
}
\end{figure*}
\begin{figure*}[hbpt]
\begin{center}
\includegraphics[width=75mm]{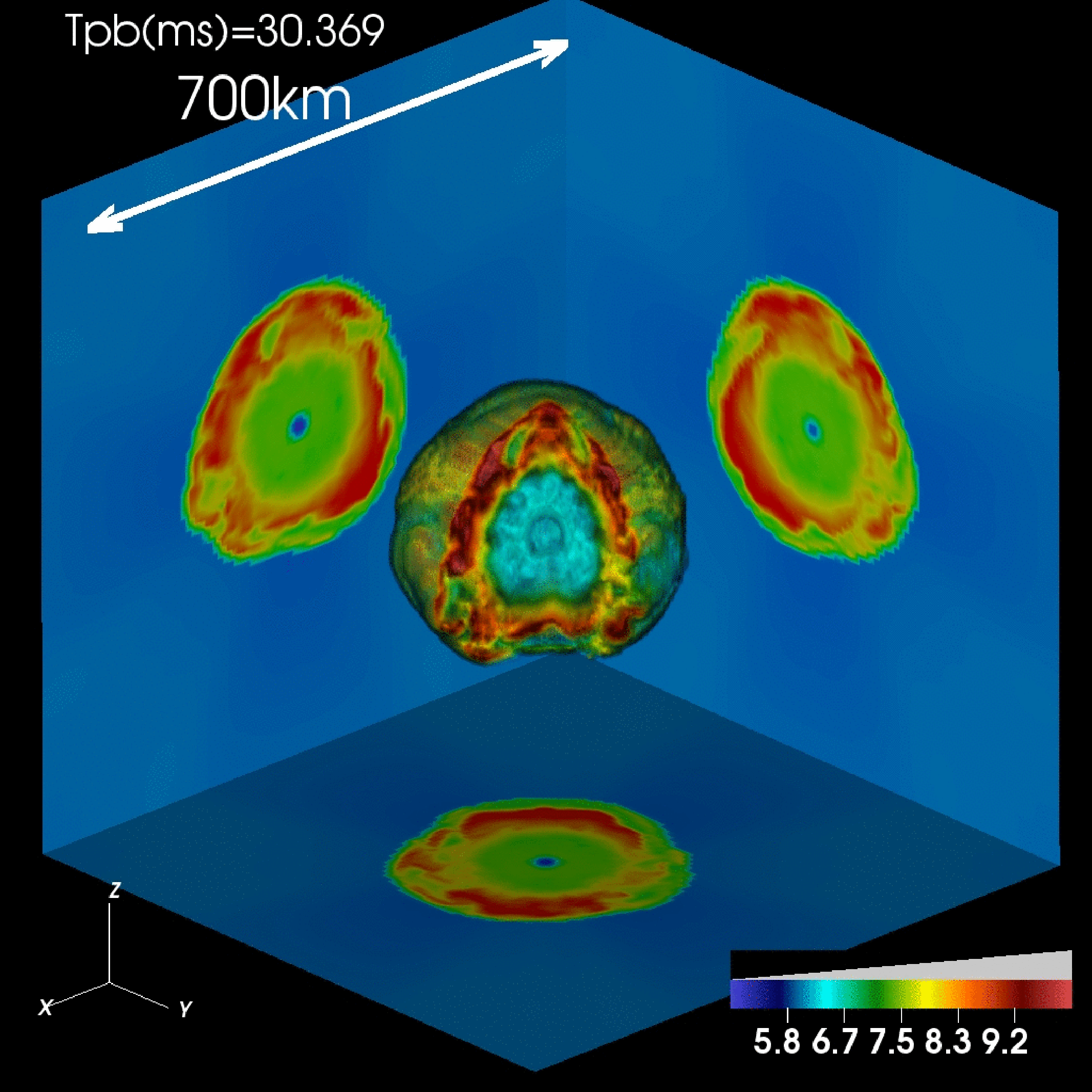}
\includegraphics[width=75mm]{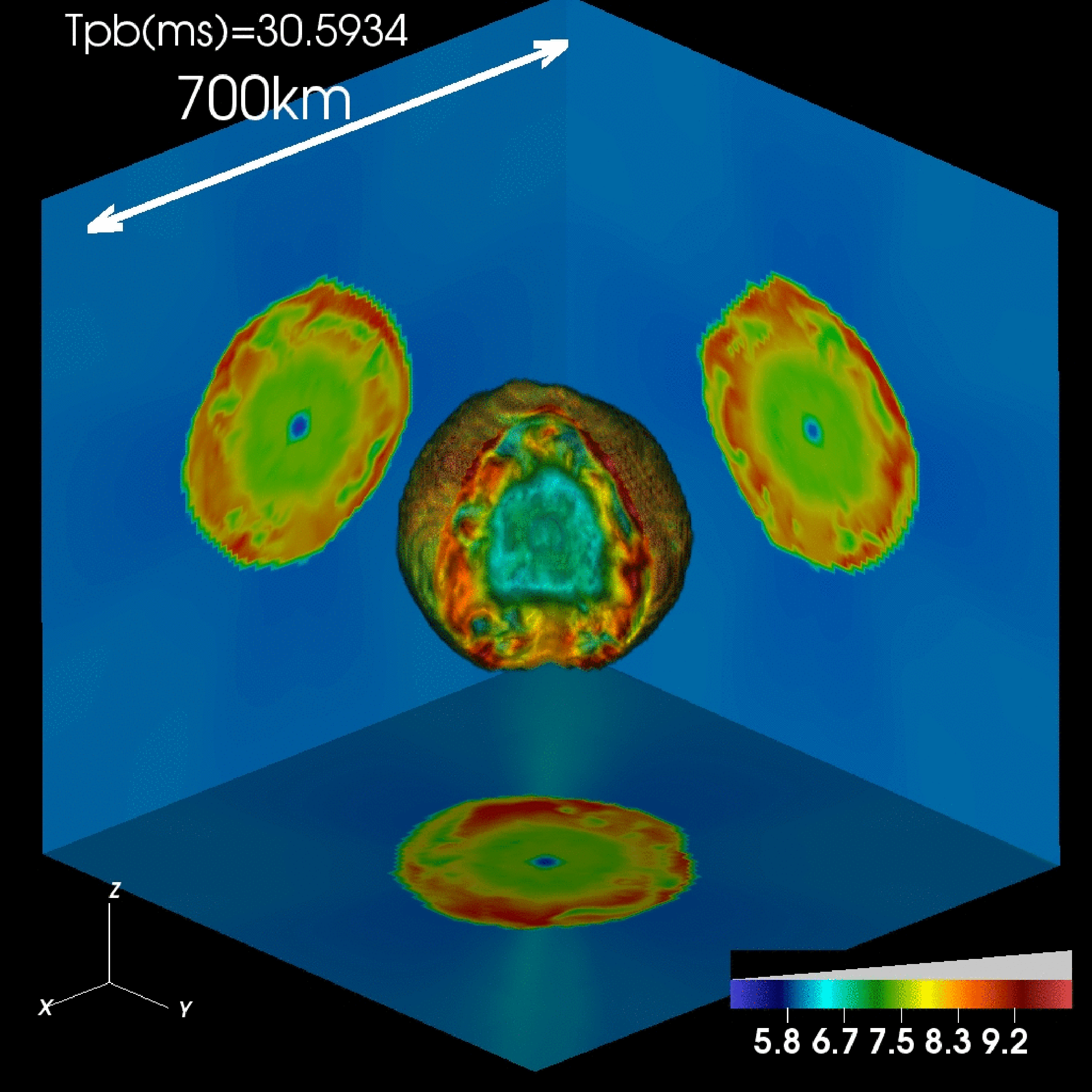}\\
\includegraphics[width=75mm]{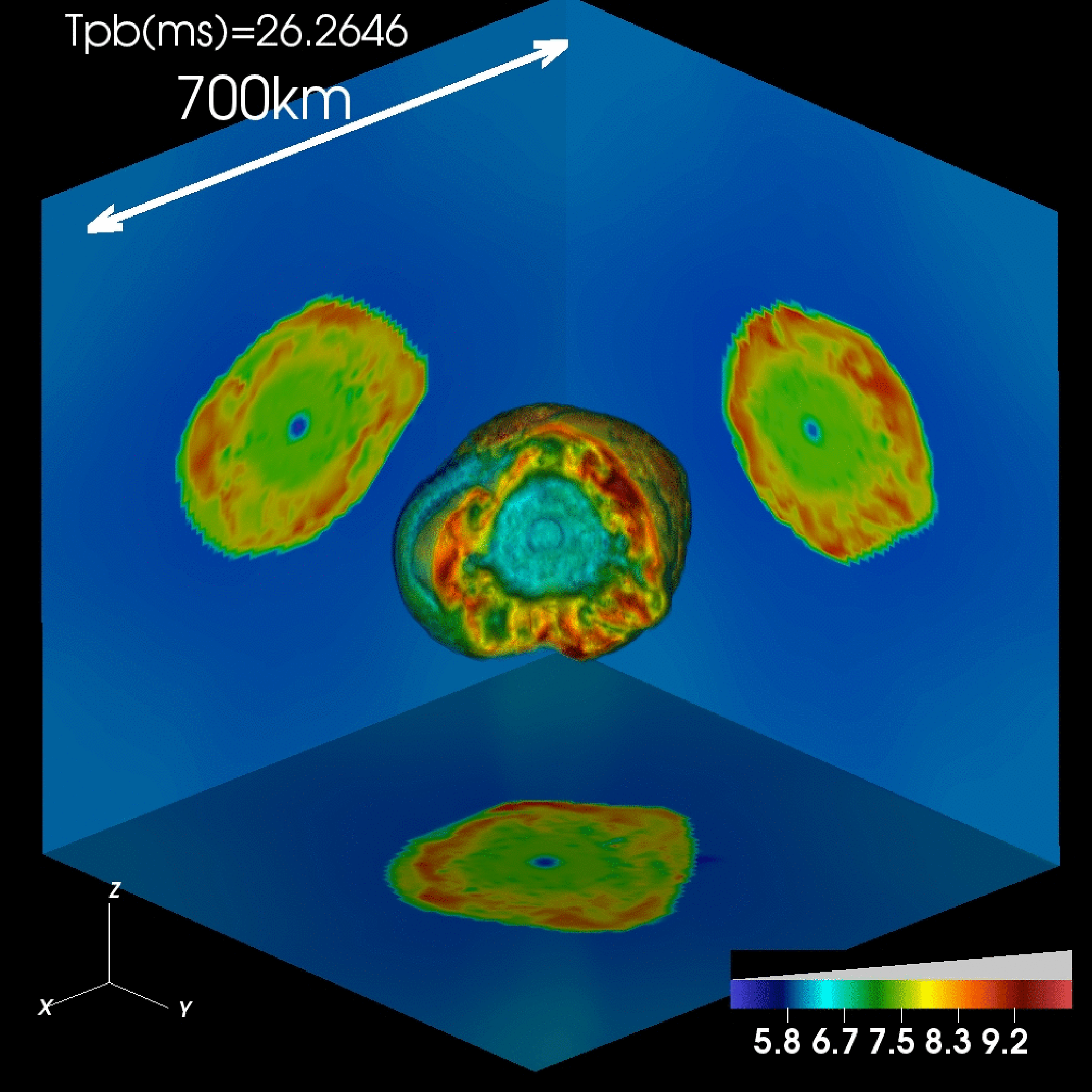}
\end{center}
\caption{\label{pic:3Dfigure_2}  Same as Fig. \ref{pic:3Dfigure_1} but for the final time snapshots
for models R0 (top left), R1 (top right), and R2 (bottom).
}
\end{figure*}
\begin{figure}[hbpt]
\begin{center}
\includegraphics[width=75mm,angle=-90.]{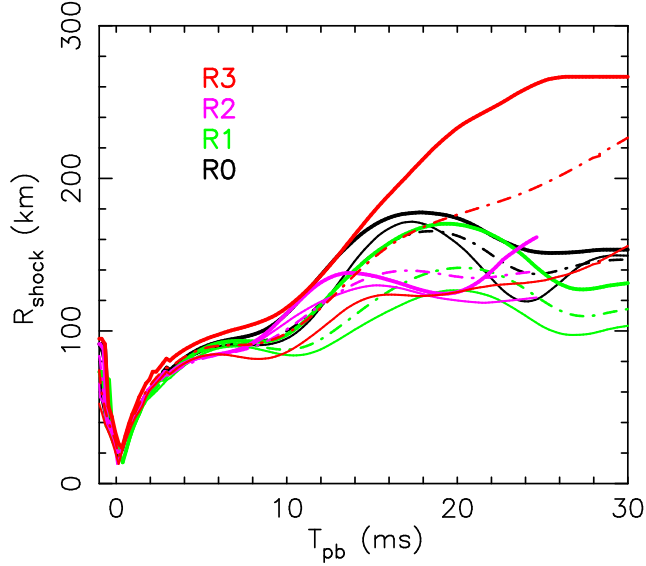}
\end{center}
\caption{\label{pic:Rshock} Evolution of angle-average (dash-dotted),
 maximum (thick) and minimum (thin) shock radii as a function of
  postbounce time ($T_{\rm pb}$).
Black, green, magenda and red lines are for models R0, R1, R2, and R3, respectively.
}
\end{figure}

Fig. \ref{pic:3Dfigure_1} shows several snapshots of 3D
 distribution of specific entropy $s(k_{\rm B}$ baryon$^{-1}$)
inside the central cube of 700km$^3$ for model R3, the most
 rapidly rotating model in this study.
   After bounce, the shock rapidly expands in the direction toward
 the equatorial plane and it then stalls at an (angle-averaged) radius
 of $\sim$260km at $T_{\rm pb}\sim$57ms.
 For the rest of models, the shapes of the shock surface
 remain nearly spherical (for model R0)
or mildly oblate (for models R1 and R2) and the average
shock radius roughly stays $\sim$150 km
within the simulation time $T_{\rm pb}\alt 30$ms (Fig.
\ref{pic:3Dfigure_2}).
These features can be also seen in Fig. \ref{pic:Rshock} which shows
 evolution of average (dash-dotted), maximum (thick), and minimum (thin)
 shock radii for all the models.
 Pushed by strong centrifugal forces, the maximum shock extent becomes
 largest for the most rapidly spinning model
  (model R3), which is consistent with Ott+12.

\begin{figure}[hpbt]
\begin{center}
\includegraphics[width=80mm,angle=-90]{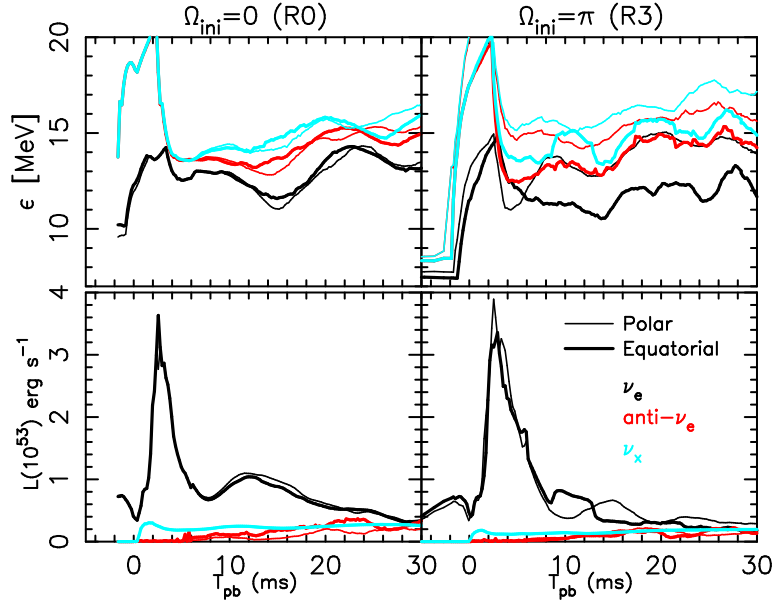}
\end{center}
\caption{\label{pic:Lnu_Enu} Postbounce evolution of the average
 neutrino energies $\varepsilon_{s_\nu,i}$
({\it top}) and the isotropic equivalent neutrino luminosities ({\it
 bottom}) for models R0 ({\it left}) and R3 ({\it right}).
Thin and thick lines are for an observer along polar ($z$ axis) and equatorial
($x$ axis) directions.
Black, red, and aqua lines represent electron, anti-electron, and heavy-lepton neutrinos, respectively.
}
\end{figure}

To see the effects of rotation on neutrino emission, we present
in Fig. \ref{pic:Lnu_Enu}
 the neutrino luminosities and the average neutrino
 energies $\varepsilon_{s_\nu,i}$. 
  Since we do not transfer the number density of neutrinos in the
  present scheme (see \citep{KurodaT12} for more details),
 $\varepsilon_{s_\nu,i}$ can be evaluated only by the following approximate way.
  We first
  project the positions of the neutrino sphere defined in the Cartesian
  grids to the 
spherical polar grids as $R_{\nu,i}(\theta,\phi)$
  for each neutrino species of $i = \nu_e, \bar{\nu}_e,\nu_x$.
Then we estimate $\varepsilon_{s_\nu,i}$ by the matter temperature
 at the neutrino sphere 
assuming that neutrinos stream freely outwards with 
possessing the information of the last scattering surface. Accordingly 
$\varepsilon_{s_\nu}$ in the spherical coordinates
$(R,\theta,\phi)$ is expressed as
\begin{eqnarray}
\label{eq:Estreaming}
\varepsilon_{s_\nu,i}(R,\theta,\phi)\equiv\varepsilon_{\nu,i}(R_{\nu,i}
(\theta,\phi),\theta,\phi).
\end{eqnarray}
Here $\varepsilon_{\nu,i}$ in the right-hand-side 
denotes the neutrino energy at $R_\nu(\theta,\phi)$, which is estimated by
\begin{equation}
\varepsilon_{\nu,i} = k_B T \frac{F_3(\eta_\nu,0)}
{F_2(\eta_\nu,0)},
\end{equation}
where $F_k(\eta_\nu,\beta)$ is the $k$-th Fermi-Dirac integral, $\eta_\nu=\mu_\nu/k_BT$ is the
degeneracy parameter with $\mu_\nu$ and $T$ representing the 
neutrino chemical potential and matter temperature, respectively on the
neutrino sphere, and $\beta$ is the relativistic factor.
The isotropic equivalent neutrino luminosity (e.g., \cite{Marek09}) is then
evaluated by $4\pi R^2 F_{(\nu)}^r$
toward arbitrary polar angles for a given radial direction.

As is expected, the neutrino luminosity and the average neutrino
energy show little
observer-angle variations for the non-rotating model R0 and
the peak $\nu_e$ luminosity ($L_{\nu_e}$)
reaches $\sim3.6\times10^{53}$ erg s$^{-1}$, 
which is quite similar to the non-rotating model in Ott+12.
 The neutrino energies of each neutrino flavor yield to the standard
 hierarchy (i.e. $\varepsilon_{\nu_e}<\varepsilon_{\bar
 \nu_e}<\varepsilon_{\nu_x}$) within our simulation time.
 As seen from the right panels of Figure 5, the neutrino luminosity and
 the average neutrino energy for model R3, on the other hand, show a clear
 directional dependence.
 The peak $\nu_e$ luminosity ($L_{\nu_e} \sim4\times 10^{53}$ erg s$^{-1}$)
 toward the polar direction is approximately 10 \% higher,
 compared to that along the equatorial direction (bottom right panel).
By comparing with the luminosity for the non-rotating model R0 (bottom
left panel), 
the luminosity for model R3 generally becomes higher (lower)
toward the polar (equatorial) direction. Due to the competition, the
total luminosity becomes
slightly smaller weaker ($\sim6$\%) for model R3 compared to the
 non-rotating counterpart.
By comparing the average neutrino energies (top left and
 right panels in Fig.5), the difference between each neutrino
 species becomes more remarkable for model R3 (i.e., the average
 neutrino energy becomes
 higher (lower) along polar (equatorial) direction, while preserving the 
 mentioned energy hierarchy).
All of these features are predominantly because of
 the rotational flattening of the core,
 by which the neutrino spheres of all the neutrino flavors are
formed more deeper inward (outward) along the polar axis (equator),
 preferentially enhancing the neutrino emission along the polar
 direction (e.g., \cite{Janka89_226,Kotake03b,Ott08}).
 In addition, the $\nu_e$ lightcurve near at the epoch of
 neutronization becomes more broader for model R3, which reflects
   the longer dynamical timescales ($t_{\rm dyn}$) for models with
   larger initial angular momentum\footnote{This is because the central
 density ($\rho_c$) supported by the centrifugal forces
 becomes smaller ($t_{\rm dyn} \propto \rho_{c}^{-1/2}$) for
   rapidly spinning models (e.g., top
  panel of Fig.1).}. These features are in good agreement with
  those obtained in Ott+12\cite{Ott12a}.

\subsection{Gravitational-Wave Signatures}
\label{sec:Emission of Gravitational Waves}

Now we are ready to discuss GW signatures in this section. After we
shortly summarize the overall waveform trend, we
perform detailed analysis on several new GW
  features that come genuinely from non-axisymmetric motions
  from the subsequent sections.
 
\begin{figure*}[t]
\begin{center}
\includegraphics[width=120mm,angle=-90]{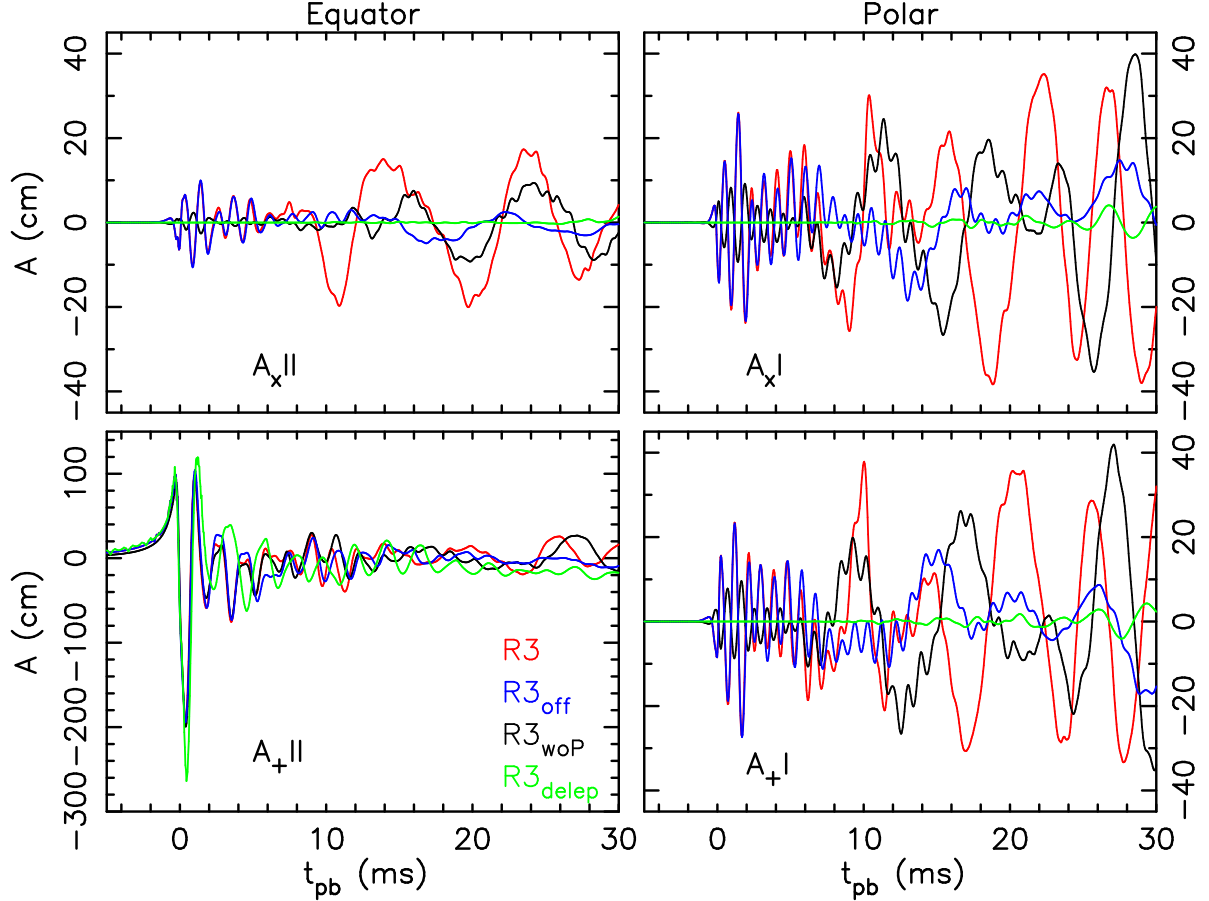}
\end{center}
\caption{\label{pic:GWAmpMat} Gravitational waveforms 
from matter motions for equatorial and polar directions (left-half and 
right-half in the pair panels, respectively) 
with two polarizations (top-half ($\times$) and 
bottom half (+)) for a series of our fastest rotating models R3
(R3 (red line), R3$_{\rm off}$ (blue line), R3$_{\rm woP}$ (black line)
 , and R3$_{\rm delep}$ (green line)).}
 \end{figure*}

The gravitational waveforms from matter motions for all the computed
models are shown in Figure \ref{pic:GWAmpMat}.\\

Note in the panel that
\begin{equation}
A_{+/\times}\rm{I}\equiv A_{+/\times}(\theta=0,\ \phi=0)
\label{equator}
\end{equation}
and
\begin{equation}
A_{+/\times}\rm{II}\equiv A_{+/\times}(\theta=\pi/2,\ \phi=0)
\label{polar}
\end{equation}
  (e.g., Eqs. (\ref{eq:A+}-\ref{eq:Ax}) and (\ref{eq:A+Nu}-\ref{eq:AxNu}))
represent the quadrupole wave amplitudes with
 two polarizations ($\times, +$) for equatorial (denoted as ``I''
 in the following) and polar (as ``II'') directions, respectively.

\begin{figure*}[t]
\begin{center}
\includegraphics[width=120mm,angle=-90]{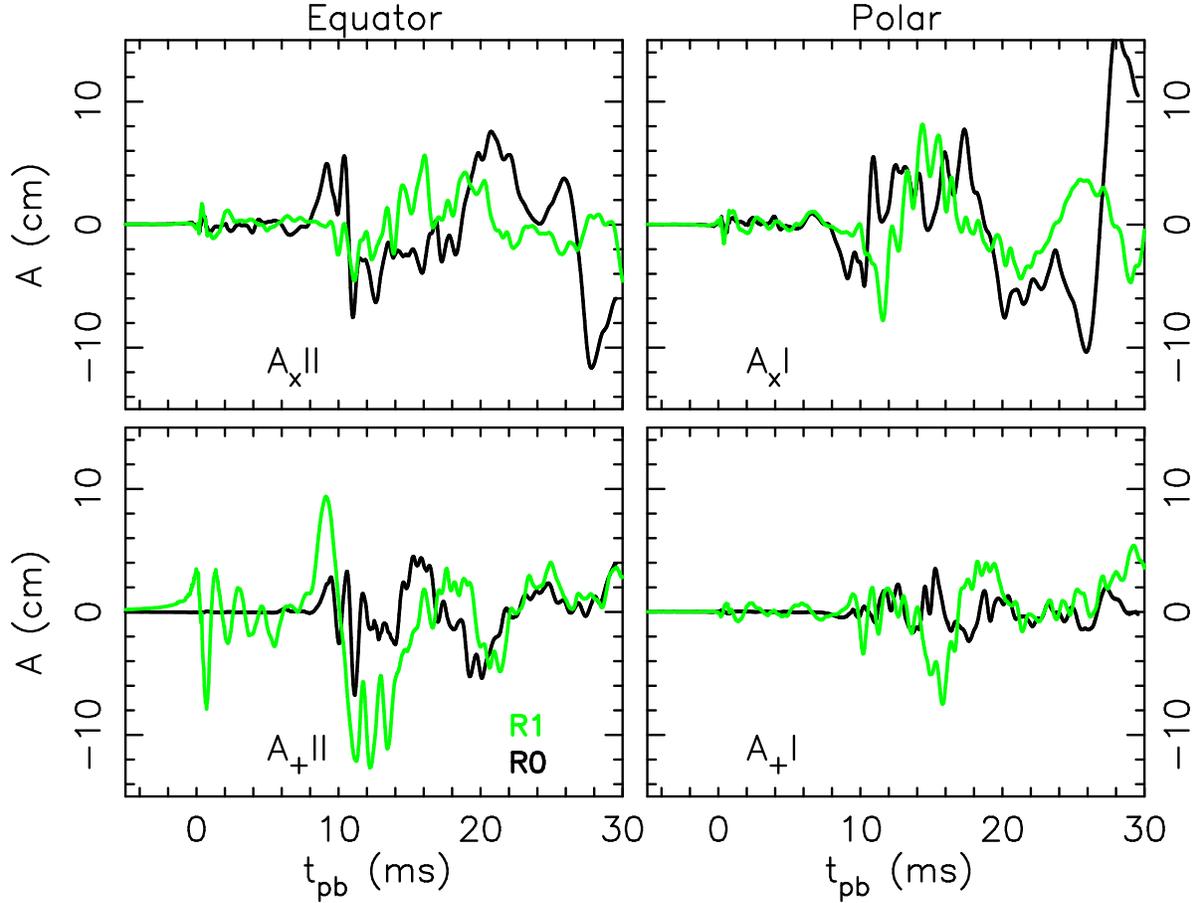}
\end{center}
\caption{\label{pic:GWAmpMat2} Same as Fig. \ref{pic:GWAmpMat}
but for models R0 (black line) and R1 (green line).}
\end{figure*}
\begin{figure*}[t]
\begin{center}
\includegraphics[width=90mm,angle=-90]{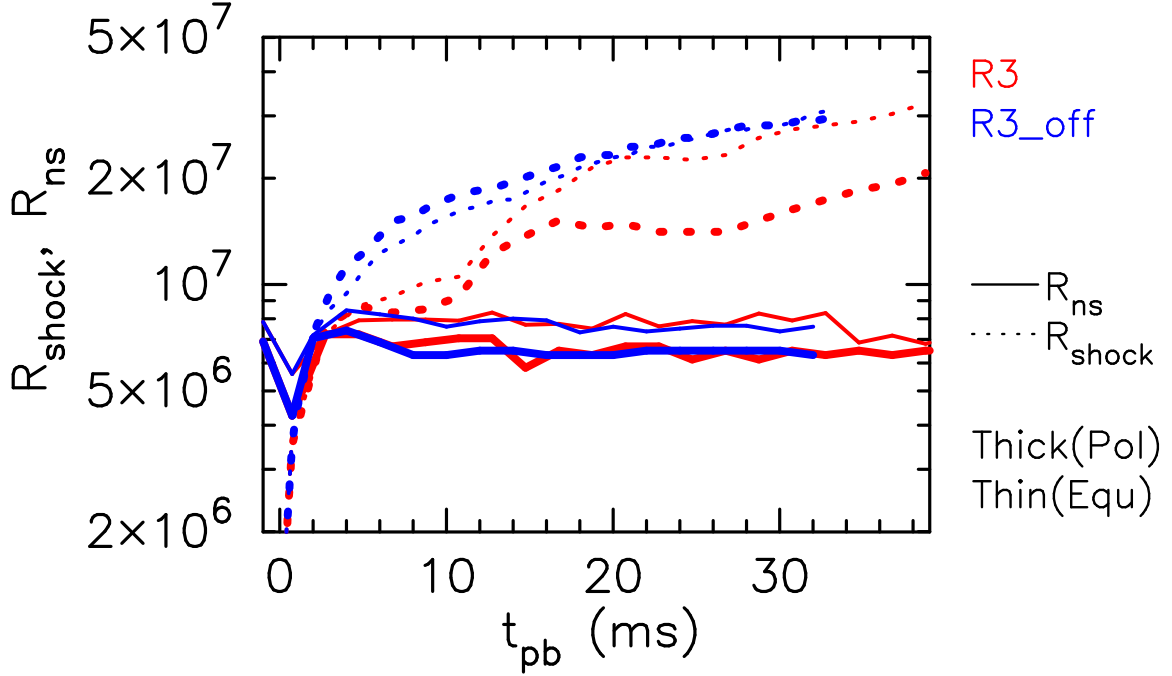}
\end{center}
\caption{\label{pic:f25.eps} Time evolution of the 
shock (R$_{\rm shock}$, dotted line)
and the PNS (R$_{\rm ns}$, solid line) radii
for models R3 (red line) and R3$_{\rm off}$ (blue line), respectively.
R$_{\rm ns}$ is defined at $\rho=10^{11}$ g cm$^{-3}$.
Thick and thin line styles represent the positions of 
R$_{\rm shock}$ and R$_{\rm ns}$
along the polar axis and equatorial plane, respectively.
}
\end{figure*}

  From the lower left panel (for model R3's),
 typical GW features of the so-called type I waveforms
 (e.g., \cite{Dimmelmeier02}) are clearly seen in the $A_+$II mode, that
  is, a first positive peak just before bounce precedes the deep
  negative signal at bounce, which is followed by the
 subsequent ring-down phase. The wave amplitude for
 model R3 is in the range of $-200\alt
 A_+{\rm II}\alt$100
 cm near bounce. This is again comparable with that in the
 counterpart model ``s12WH07j3'' in Ott+12. {\bf From Ott+07 who employed
the same Shen EOS (but with a more simplified treatment for
 deleptonization), their counterpart
  model (the $\beta$  parameter $\sim 6.7\%$ at bounce) has
 $-240\alt A_+{\rm II}\alt$90 cm, which is also in good agreement with
 our model R3$_{\rm delep}$. Regarding the non-axisymmetric GW 
(green lines in the right panels) in the first 10 ms postbounce,
 the amplitude stays negligibly small (the maximum amplitude is at most $\sim10^{-4}$cm), 
and the overall features of the waveforms during our simulation time 
($\sim 30$ ms postbounce)
 are consistent with those in Ott+07 (their model E20A)}. The wave amplitudes for our non-rotating and slow-rotating 
 models stay much smaller ($A_+{\rm II}\alt$10 cm) during the early postbounce evolution (models R0 (black line) and R1 (orange line) 
in the right pair panels of Fig. 6).

By comparing model R3 (red line) with R3$_{\rm off}$ (blue line),
 deviation of the two waveforms becomes remarkable
 only after $\sim8$ ms after bounce when the neutronization ceases.
 The GW amplitudes become higher for model with deleptonization (R3)
 compared to the counterpart model (R3$_{\rm off}$) for which 
deleptonization effects are turned off manually after bounce.
This is also consistent with recent report by Ott+12.
As already pointed out by \cite{Scheidegger10}, this is because
 neutrino cooling in the postbounce phase leads to a more compact 
 core (e.g., Fig. \ref{pic:f25.eps}) with bigger enclosed mass inside, 
which results in 
 a more efficient GW emission. In Fig. 8, it can be also shown
 that due to deleptonization, the maximum shock extent is 
 smaller for model R3, which makes convectively unstable regions
(R$_{\rm ns}\alt$R$\alt$R$_{\rm shock}$) more compact than
 for model R3$_{\rm off}$. 
Then the growth time scale of prompt convection tends to be 
shorter which leads to strong inhomogeneity and emissions of 
more powerful GWs toward the spin axis together with rotation.
 These results confirm the previous findings 
(e.g., \cite{Scheidegger10,Ott07_prl}) that accurate 
neutrino transport is required for a reliable GW prediction. 
In this respect the prediction power of our approximate neutrino 
transport scheme is limited, which should be tested by 3D GR models 
 with more detailed neutrino transport.

\begin{figure*}[t]
\begin{center}
\includegraphics[width=100mm,angle=-90]{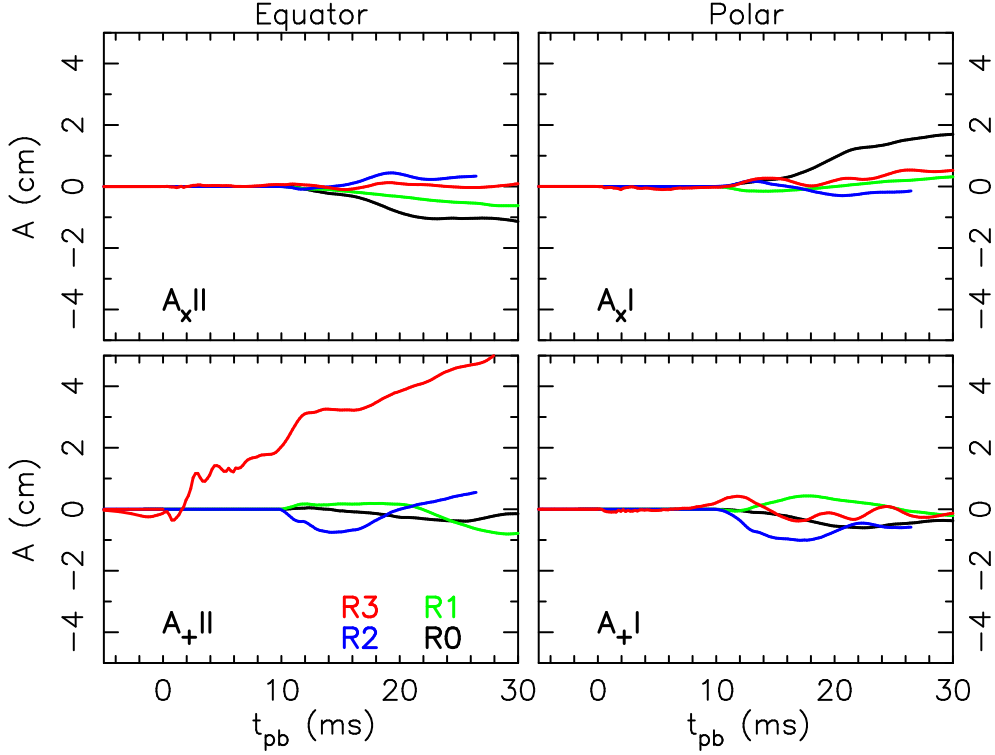}
\end{center}
\caption{\label{pic:GWAmpNu} Same as Figure 6 but for 
 the waveforms from anisotropic neutrino emission (sum of
 the contributions from all the neutrino species).}
\end{figure*}

Fig.\ref{pic:GWAmpNu} shows the waveforms from
 anisotropic neutrino emission (Eqns.(\ref{eq:A+Nu},\ref{eq:AxNu})).
Due to the {\it memory effect} inherent to the neutrino GWs (e.g.,
 \cite{Braginskii87}), the
 waveforms exhibit much slower temporal variation compared to the matter GWs.
 From the lower left panel, the largest amplitude of the neutrino GWs
 is obtained for the $A_+$II mode in model
 R3 (red line), which reaches $\sim6$ cm with a
 quasi-monotonically increasing trend during the simulation time. This
 characteristic feature was already reported so far either in
 2D simulations employing detailed neutrino transport 
 \cite{EMuller04} or in 3D simulations with idealized setups \cite{Kotake11}. This comes from the (mentioned) stronger
 neutrino emission along the rotation axis due to the deformed neutrino
 sphere in the rapidly spinning cores. Except for model R3,
  the wave amplitudes from neutrinos stay around a few cm
 during our short simulation time.

\subsection{Non-axisymmetric imprints}\label{nonaxis}
Now let us have a look at again the matter GW signals for
the $A_{+/\times}$I modes (emitted toward the pole,
e.g., the right-half panel of Figure 6 ({\it
left})). Note that these signals come genuinely from
non-axisymmetric matter motions. The red lines (model R3) in these panels
 show an oscillating behavior with two different modulation timescales
 (:$\tau_{\rm mod}$), firstly
 in a very short timescale ($\tau_{\rm mod} \lesssim 1$ ms)
 promptly after bounce within $t_{\rm pb} \lesssim 8$ms, and secondly in
 a relatively longer modulation ($\tau_{\rm mod} \gtrsim 10$
 ms) after that. These features are only clearly visible for our
rapidly rotating model R3.

 For model R3, the wave amplitudes in the first epoch
($t_{\rm pb} \lesssim 8$ms) reach $|A_{+/\times}$I$|\sim$20 cm,
while non- and slowly rotating
 models (R0 and R1) meanwhile produce very little GW emission ($|A$I$|\alt1$ cm).
 By comparing with the luminosity curves in Fig.5, this epoch is shown
  to closely correlate with the neutronization phase. During this epoch,
    the prompt shock propagates rapidly outward with capturing electrons
  and dissociating infalling iron group nuclei until the prompt shock
  stalls at around $t_{\rm pb}\sim10$ ms. As seen from Fig.4, the shape
   of the shock for all the models is rather spherical at this time,
   regardless of the
   difference in the initial rotation rates. 
In such a short duration after bounce,
   possible reason of producing the
 non-axisymmetry is less likely to be the low-$T/|W|$ instability,
 not to mention the SASI. 

{\bf The black and red curve in Figure 10 corresponds to the $A_+$I and 
$A_{\times}$I 
mode waveform of model R3 (see also red line in Figure 6), respectively.
At the very early postbounce phase ($t_{\rm pb} \lesssim 8$ ms),
 the phase of the quasi-oscillatory pattern of the two modes (black
 and red curves) is shifted about $\pi/2$. Such feature of the 
 phase-shifted pattern should be coincided with the bar mode deformation ($\ell=2, m=2$).\footnote{as it is well known in the case of the GW emission from 
binary coalescence.} The (non-dimensional) amplitudes of the $m = 2$ 
mode in Figure 10 (pink line) supports this anticipation. Here 
we estimate the normalized azimuthal Fourier components $A_m$ 
as
\begin{eqnarray}
\label{eq:Am}
A_{\rm m}=\frac{\int_0^{2\pi}\rho(\varpi,\phi,z=0)e^{i{\rm m}\phi}d\phi}{\int_0^{2\pi}\rho(\varpi,\phi,z=0)d\phi},
\end{eqnarray}
where the $m=2$ mode is evaluated at a given radius of $\varpi\equiv\sqrt{x^2+y^2}=20$ km. Regarding the mode amplitudes and the GW emission, the $m = 2$ mode 
amplitudes is $\sim 10^{-2}$ (e.g., the label of $A_m$ in Figure 10)
at the very early postbounce phase, which results in the GW emission 
of $\sim$ 20 cm. This is in good agreement with \cite{Ott07_prl}
 (in their Figure 3) who showed the $m = 2$ mode amplitudes of 
$10^{-2}$ leads to the wave amplitude $\sim$ 30 cm. These 
results might suggest the $m=2$ mode from the seed perturbations 
could dominantly act as the source of the GW emission.
But it should be noted that our model 
R$3_{\rm woP}$ that does not have initial seed perturbations 
(see black line in Figure 6) produce non-negligible GW emission 
at the very early postbounce phase ($\lesssim 6$ cm), although the 
 wave amplitude is smaller than that for model R3 that has seed 
initial perturbations (Figure 12)\footnote{The non-vanishing components could 
come from intrinsic numerical perturbations, which are unavoidable for 
any code using a Cartesian grid (see discussions in \cite{Ott12b}.)}.
In addition, we have to add that 
when we employ only the Liebendoerfer's deleptonization scheme in 
model R3 (without neutrino leakage scheme), the wave amplitude at the 
very early postbounce phase becomes very small ($\sim 10^{-4}$ cm). 
Remembering that the leakage scheme is nothing but 
a very crude approximation of neutrino transport, further 
investigation is needed to clarify the in-depth analysis about the 
impacts of initial seed perturbations on the early postbounce GW signals.
 This should require a systematic study, in which a variety of 
3D-GR models are to be computed with refined numerical resolutions and 
with more sophisticated transport scheme, which we consider as a very 
important extension of this study.}
\begin{figure*}[t]
\begin{center}
\includegraphics[width=60mm,angle=-90]{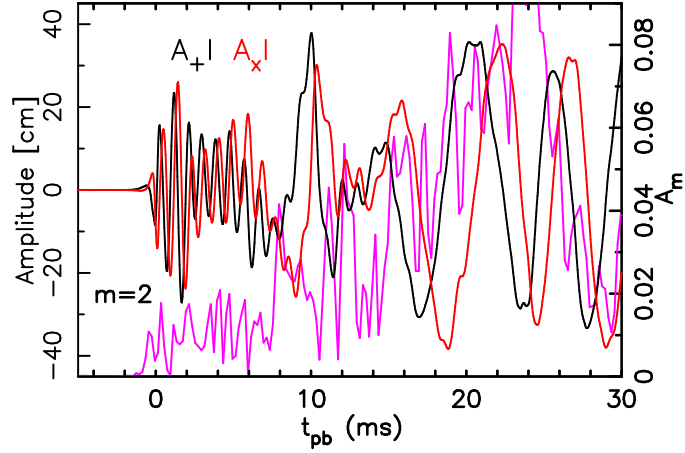}
\end{center}
\caption{\label{pic:Phase_Shift} Gravitational waveforms 
from matter motions for polar direction in model R3.
Black and red lines are for $A_{+}$I and $A_{\times}$I, respectively.
The pink curve represents non-axisymmetric mode amplitude with $m=2$ 
(e.g., Eq.(\ref{eq:Am})) of 
matter distribution at a given radius $\varpi=20$ km.}
 \end{figure*}


\begin{figure}[t]
\begin{center}
\includegraphics[width=70mm,angle=-90]{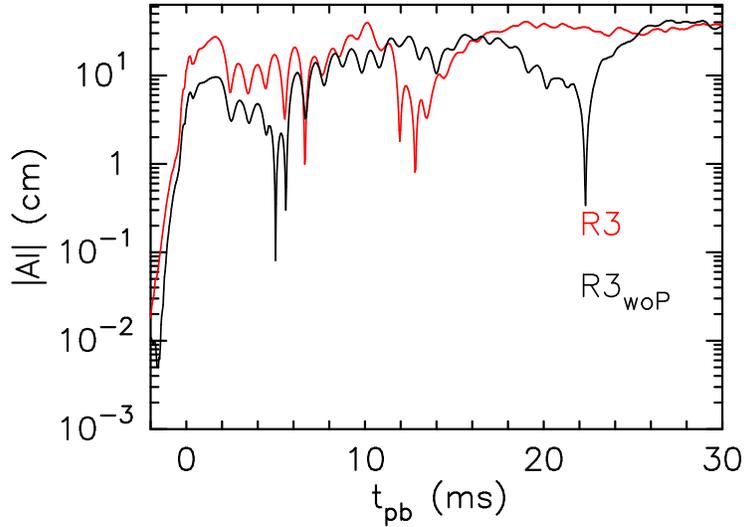}
\end{center}
\caption{\label{pic:GWAmp_perturb} The absolute GW amplitude of $|A\rm
 I|$ (toward the polar direction,
 Eq.(\ref{polar})) for our most rapidly rotating
 model R3 with (red line) and without (black line) initial seed perturbations.}
\end{figure}

As the neutronization phase comes to an end at around $t_{\rm pb}\sim 8$ ms
(Fig.5), the prompt shock simultaneously stagnates, triggering
the entropy-driven prompt convection behind.
 After that, the waveforms for moderately and rapidly spinning models
(R2 and R3) exhibit a longer modulation with slightly larger amplitudes
  ($|A|\sim20$ cm) compared to those near the neutronization phase.
 Non- (R0) and slowly (R1) rotating
models emit roughly $\sim10$ times stronger GWs.
These amplitudes during prompt convection are in accord with previous studies,
$|A|\sim15$ cm in \cite{Ott12a,Ott12b}, $|A|\sim20$ cm in \cite{BMuller13} and $|A|\sim8$ cm in \cite{Scheidegger10}.

\begin{figure*}[htpb]
\includegraphics[width=80mm,angle=-90]{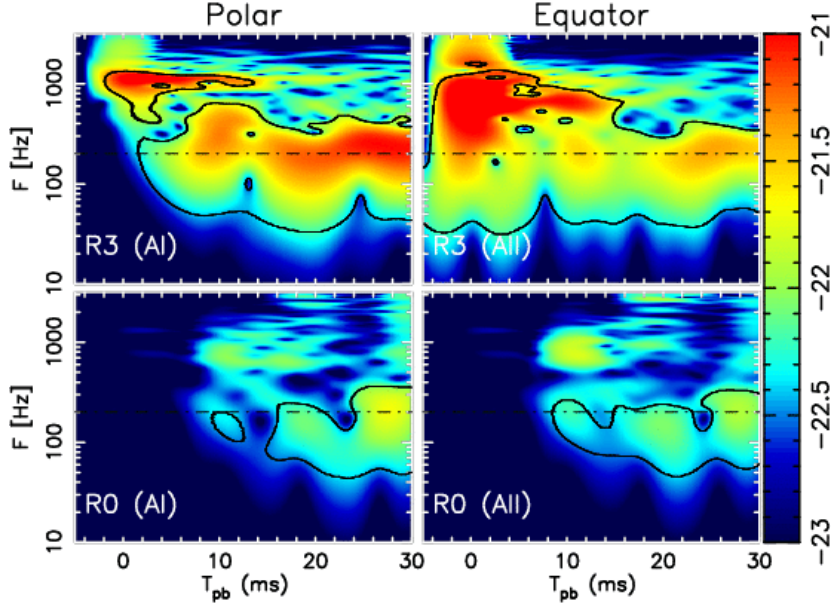}
\caption{\label{pic:SpecTime_R0R3} Color-coded GW spectrograms ($\log_{10}
  (h_{\rm char})$, Eq. (\ref{hchar})) for models R3({\it top}) and
  R0({\it bottom}).
Left and right panels are for polar $(\theta,\phi)=(0,0)$ and equatorial
 $(\theta,\phi)=(\pi/2,0)$ directions, respectively.
  The model name with the polarized mode is given in the lower left
  corner of each panel (for example, ``R3 (AI)'' in the top left panel).
 The contour drawn by black solid line in each panel corresponds to a threshold
   beyond which the signal-to-noise ratio (SNR)
  for KAGRA \cite{KurodaK10} exceeds unity (for a Galactic event). To
  guide the eyes (see text for details), the horizontal dotted line
  (200 Hz) is
  plotted. To clearly present the postbounce GW signatures, we set the
  maximum value of the color scale as ${\rm
  log}_{10}(h_{\rm char})=-21$, while the wave amplitude is actually more
  higher near bounce
  (${\rm log}_{10}(h_{\rm char})\sim-20.3$ in the AII mode (top right panel)).}
 \end{figure*}

As seen from red line in Fig.6 (left panel),
the waveform of model R3 has a clear sinusoidal modulation,
 which possesses a $\pi/2$ phase shift between $A_{+}{\rm I}$ and
 $A_{\times}{\rm I}$. This feature persists until the end of simulations
($t_{\rm pb}\alt 50$ ms in this study).
To understand the origin of the sinusoidal signature,
we present a spectrogram analysis in Fig. \ref{pic:SpecTime_R0R3}.
In the figure, the angle-dependent
characteristic strain $h_{\rm char}$

 \begin{eqnarray}
h_{\rm
 char}(\theta,\phi,F)=\sqrt{\frac{2}{\pi^2}\frac{G}{c^3}\frac{1}{D^2}\frac{dE(\theta,\phi)}{dF}},
 \label{hchar}
 \end{eqnarray}
 is plotted (e.g.,
\cite{Flanagan98,Ott04,Murphy09}),
where $D$ represents the source distance that we assume as $D=10$ kpc
 (unless otherwise stated), $F$ is the GW frequency,
 and $dE(\theta,\phi)/dF$ is the GW spectral energy density.
 $dE(\theta,\phi)/dF$ is given by\footnote{The angle dependence
 $(\theta,\phi)$ is omitted for the sake of brevity.}
\begin{eqnarray}
\frac{dE}{dF}=\frac{\pi}{4}\frac{c^3}{G}F^2
\left(|\tilde A_+(F)|^2+|\tilde A_\times(F)|^2\right).     
\end{eqnarray}
Here $\tilde A(F)$ denotes the Fourier component of $A(t)$
\citep{Murphy09} that is defined as 
\begin{eqnarray}
\label{eq:Atilde}
\tilde A_{+/\times}(F)=\int_{-\infty}^\infty A_{+/\times}(t) H(t-\tau) e^{-2\pi i Ft}dt,
\end{eqnarray}
 where the width of the Hann window $\tau$ is set as 10 ms.

\begin{figure*}[htpb]
\includegraphics[width=80mm,angle=-90]{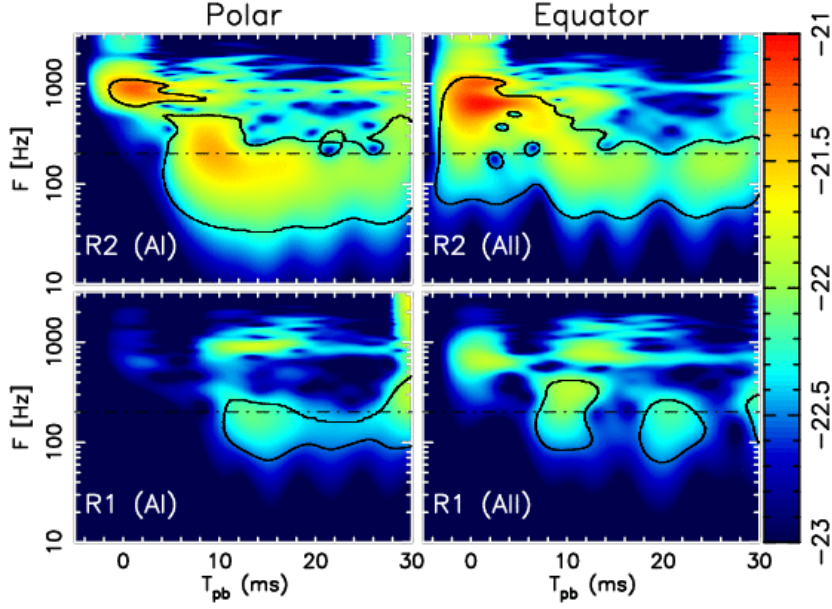} 
\caption{\label{pic:SpecTime_R1R2}Same as Fig. \ref{pic:SpecTime_R0R3}
 but for models R2 ({\it top}) and R1 ({\it bottom}).}
\end{figure*}
The spectrogram for the equatorial GW (top right panel) near bounce
 ($t_{\rm pb}\sim 0$ ms) shows a power excess (colored by yellow to
 reddish regions) in the range between $100$ $\alt F\alt1000$ Hz, which
   is associated with the best-studied type I signals (e.g.,
 \cite{Dimmelmeier07,Ott07_prl,Ott07_cqg})\footnote{In line with previous works,
these type I signals are likely within the
detection limits of the next generation detectors for a Galactic source,
 when the $\beta$ parameter at bounce exceeds $\sim1\%$ (e.g., model R2 in this work).}.
 Not surprisingly, this feature is hardly seen in the non-rotating model (bottom right
 panel in Fig.8). Since the deviation from spherical symmetry for model
 R0 is very small
  in the early postbounce phase, little angular variations are seen in the GW
   spectrogram (compare bottom left with bottom right
 panel). On the other hand, the spectrograms for model R3 have
 much clearer angular dependencies (compare top left with top right
 panel). The polar GW spectrogram (top left panel) has two distinct
  power-excess {\it islands} that are enclosed by black solid line (satisfying
 the signal-to-noise ratio: SNR$\agt1$ for a Galactic source by
  the second generation detectors). The typical GW frequency of the first island is about
 $\sim$ 1000 Hz between $0 \lesssim t_{\rm pb} \lesssim 10$ ms
 (a reddish zone near the top left corner in the top left panel), and it
 is around $200-250$ Hz for the second island with its excess clearly
   visible from $t_{\rm pb} \gtrsim 10$ ms (e.g., a horizontal reddish
 stripe covering its peak around $200-250$ Hz in the panel).

  The first island, which is narrower (in width and height)
  than the second one in the
  frequency-time domain, originates from the initial seed perturbations
  that we mentioned above. It is interesting to point out that the
   resulting GW amplitudes satisfy SNR$\agt1$ (enclosed by black solid
  line) for a Galactic source. In addition to the well-studied
  equatorial GWs\footnote{We shortly call GWs seen from
  the polar (equatorial) direction as polar (equatorial) GWs, respectively.} (associated with the type I signals), our results suggest
 that the {\it polar} GWs just after bounce have also a unique signature, which is
  produced by the percent levels of the precollapse density
  fluctuations seeded in the rapidly rotating cores.

  Regarding the second island, the characteristic frequency $F_{\rm
  char}=200\sim 250$
  Hz for model R3 ($t_{\rm pb}\agt 10$ ms, top left panel)\footnote{e.g.,
  the horizontal dotted line that closely divides the island into two.} is higher
  than those ($F_{\rm char}\sim100$ Hz) in models R0 and R1
  (compare Figs. 8 and \ref{pic:SpecTime_R1R2}).
Note that the lower characteristic frequency ($F_{\rm char}\sim100$ Hz)
observed in our non-rotating and
  only mildly rotating models is
  consistent with recent results by \cite{BMuller13} who performed 2D
  GR simulations including detailed neutrino transport.
According to their analysis, the GWs during prompt convection are
predominantly generated  by radially propagating acoustic waves above the PNS.
 With the typical sound velocity there ($C_{\rm s}\sim10^9$ cm s$^{-1}$)
 and the shock radius ($R_{\rm shock}\sim100$ km), the
  typical frequency in our 3D-full-GR results ($F_{\rm char} \sim
  C_{\rm s}/R_{\rm shock}\sim100$ Hz) can be also reasonably
   estimated. On the other hand, the characteristic frequency for
  the second island ($F_{\rm char} = 200\sim 250$ Hz, most clearly visible for
  model R3, e.g., top left panel in Fig.8) is shown to be systematically
 higher than the GW component solely from the propagating acoustic waves.
From the next section, we look into the reason of the
   higher frequency in more detail. Touching on the detectability,
  the relevant GW frequencies (regardless of the difference)
  are in the range of $F_{\rm char}
  \sim100-250$ Hz, which are close to
  the maximum detector sensitivity for the second generation interferometers
  (e.g., KAGRA and Advanced LIGO \cite{Harry10,somiya}).

\subsection{GW Emission from One-Armed Spiral Waves}
\label{sec:GW Emission from One-Armed Spiral Wave}
In this section, we are going to discuss
 that the characteristic frequency
($F_{\rm char}=200\sim 250$ Hz); clearly visible
in the GW spectrogram for model R3 has a
tight correlation with
the one-armed spiral waves.
   We will also show
  that these features can be naturally explained by the acoustic feedback between the stalled shock and the
  rotating PNS surfaces, which is thus reconciled with the
    well-established picture of the SASI (\cite{Foglizzo07,Foglizzo09,Fernandez10}).
  
\begin{figure}[htbp]
\begin{center}
\includegraphics[width=95mm,angle=-90]{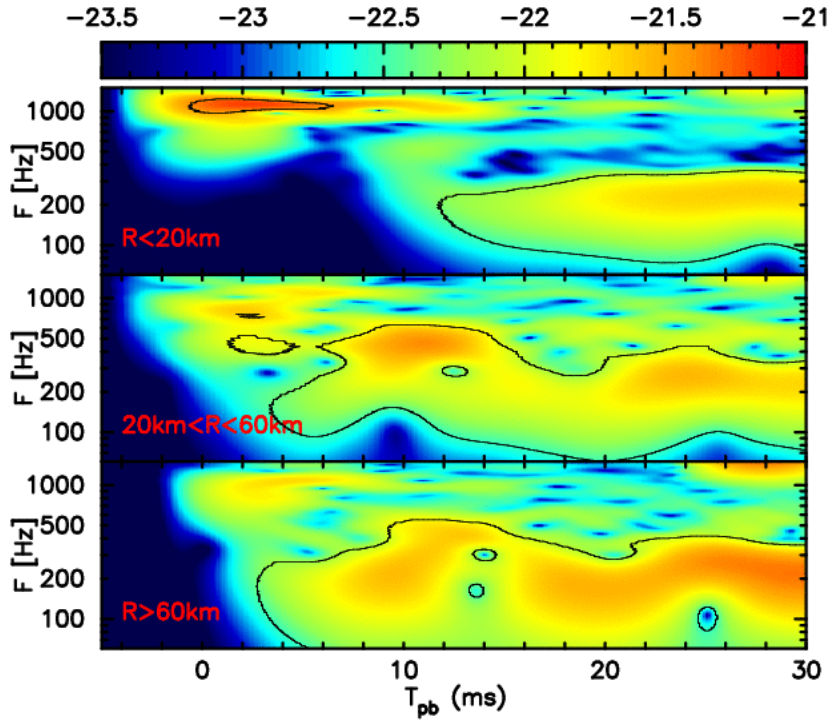}
\end{center}
\caption{\label{pic:dEdF_radial} Similar to the right panels
 of Fig.\ref{pic:SpecTime_R0R3}, but contributions to
the first time derivative of the mass quadrupole moment $\dot I_{ij}$
 are presented for different radial locations from the innermost ($R<20$km, {\it top}), 20km$<R<60$km ({\it middle}),
 to the outer region ($R>60$km, {\it bottom}), respectively. Note in
 this panel that GWs seen from polar direction are plotted.}
\end{figure}

Similar to the right panels in Fig.\ref{pic:SpecTime_R0R3}, but we plot in Fig. \ref{pic:dEdF_radial} 
  contributions to the first time derivative of the mass
  quadrupole moment (Eq.(\ref{eq:dIijdt})) from different radial
  locations (seen from the polar direction).
The top, middle, and bottom panels represent $h_{\rm char}$
 evaluated by the spatial integral in the following range,
 within $R<20$ km, $20<R<60$ km, and $R>60$ km, respectively.
It can be seen that the GW emission in the second island (again, the horizontal ($F_{\rm char}=200\sim 250$ Hz) reddish zone 
after around $T_{\rm pb} \sim 10$ ms) are radiated mainly
from above the PNS ($R\agt60$ km, bottom panel) during prompt convection.
Note that the spatial location itself is similar to the one reported
in \cite{BMuller13} for their non-rotating model.

\begin{figure*}[htpb]
\begin{center}
\includegraphics[width=74mm,angle=-90]{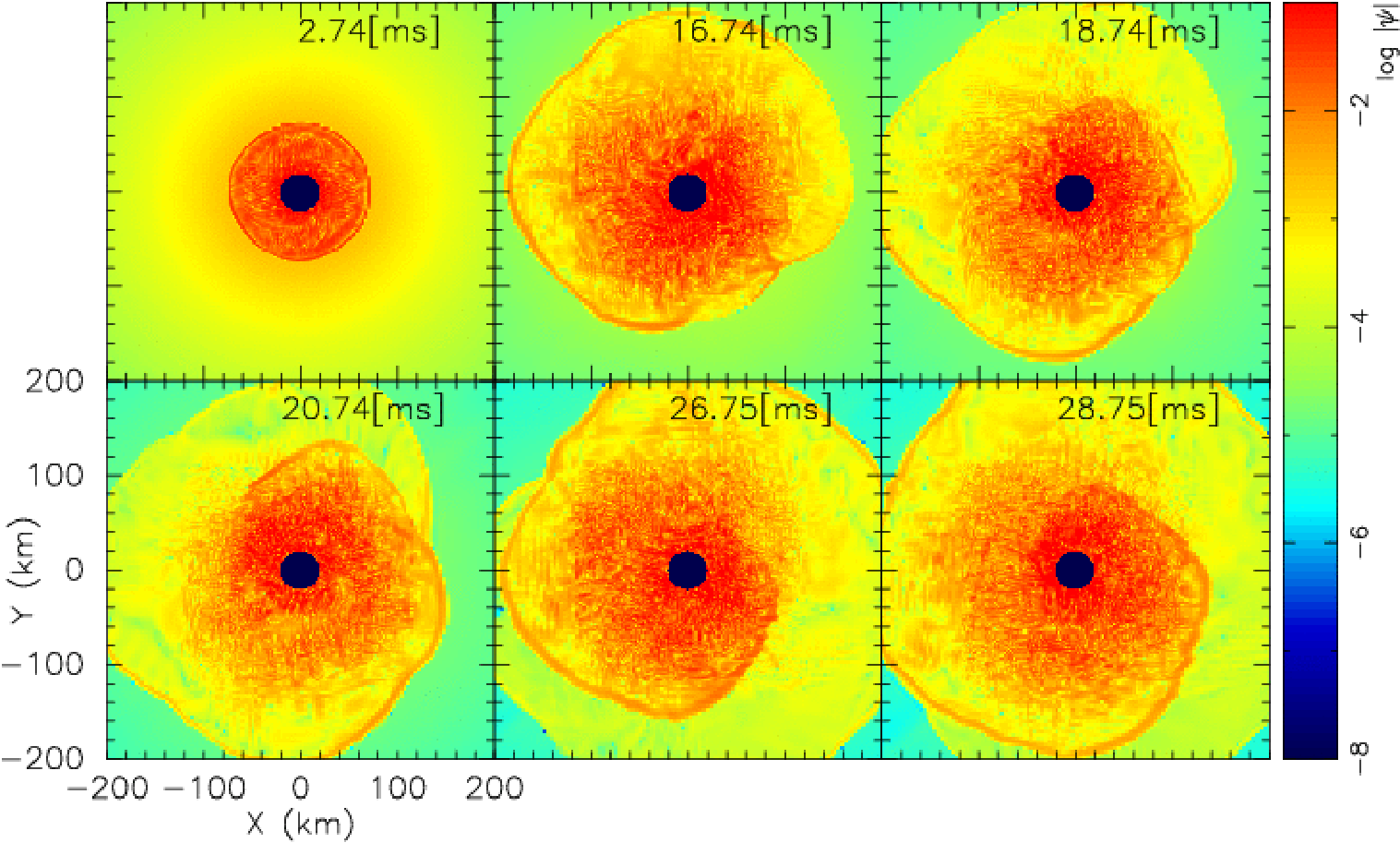}
\includegraphics[width=74mm,angle=-90]{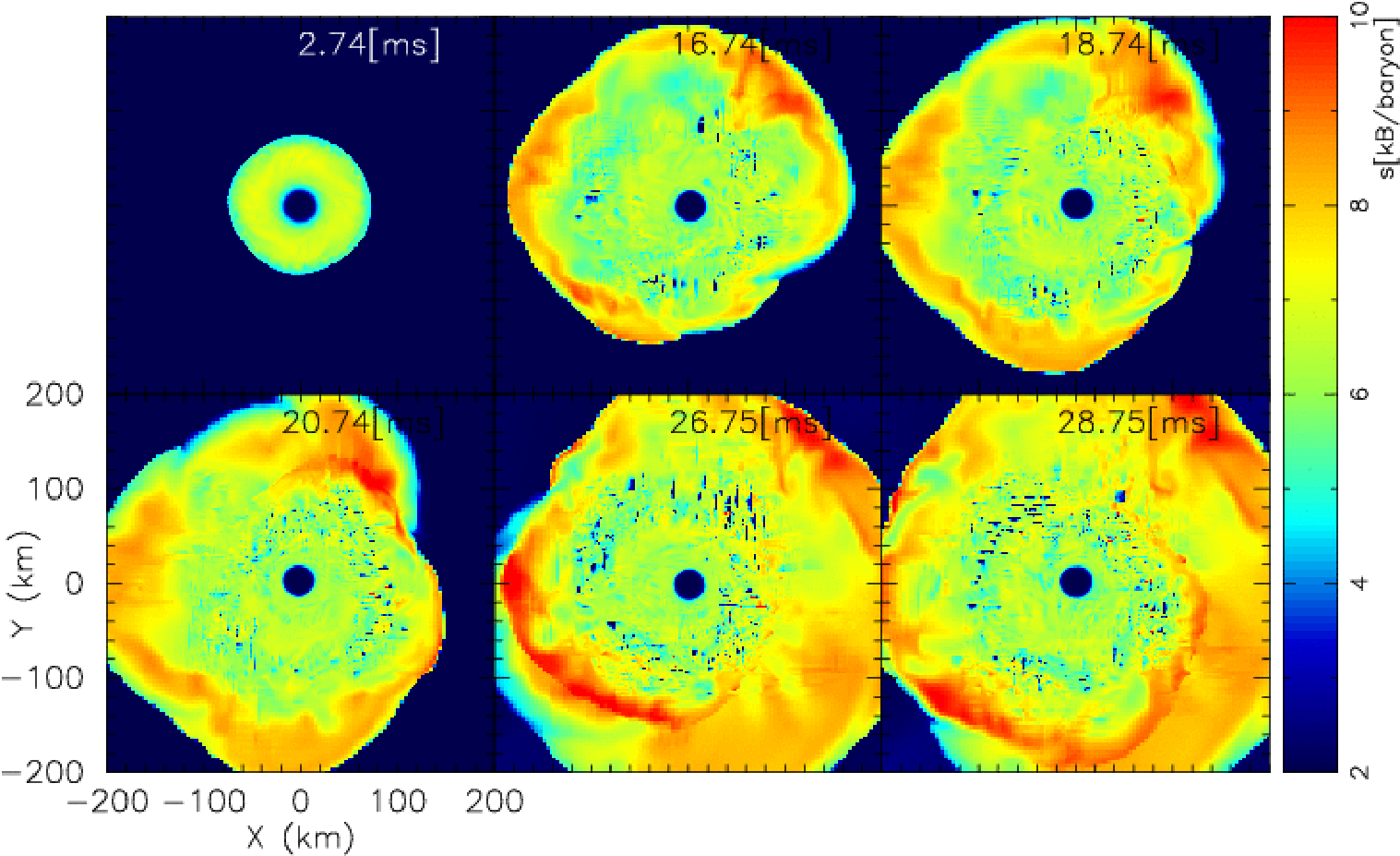}
\end{center}
\caption{\label{pic:GW_loc} Colormaps of the integrand $\psi$ in the
  quadrupole formula ({\it top})
and entropy (per baryon, {\it bottom}) on the equatorial plane
 at selected time slices (denoted in the top right corner in each mini
 panel) for model R3.
In the upper panels plotting $\psi$, the central region $R\le20$ km is excised
 to show a clear contrast. Note that the color scale is in a
 logarithmic scale and it is normalized by
  the maximum value of $\phi$.}
\end{figure*}

To see how the source of the strong GW emitter evolves with time,
 Fig.\ref{pic:GW_loc} shows the integrand $\psi$ in the quadrupole formula
 (top panel);
 \begin{eqnarray}
\psi\equiv(A_+{\rm I})^2+(A_\times {\rm I})^2= (\ddot I_{xx}-\ddot I_{yy})^2+(2\ddot I_{xy})^2.
\end{eqnarray}
 and the specific entropy (bottom panel)
 at selected time slices.  
From the upper panels (regarding $\psi$), it can be seen that
the one-armed spiral wave starts to form at
around (e.g., $T_{\rm pb} = 16.74$ ms, top middle panel ({\it upper}))
for model R3, and then it keeps rotating in a counter-clockwise manner
(compare $T_{\rm pb} =
 26.75$ ms and $T_{\rm pb} = 28.75$ ms) until the end of simulations.
 From the two panels (at $T_{\rm pb}=26.75$ and 28.75 ms),
  the spiral wave
 rotates around $\sim90^\circ$ during 2 ms, thus the rotational period (frequency)
 can be estimated as $T_{\rm rot}\sim8$ ms and $2/T_{\rm rot} \sim 250$ Hz\footnote{
Here 2 in the numerator comes from two polarized wave modes
($+/\times$).}. Note that this is close to the mentioned GW frequency
  $F_{\rm char}= 200\sim 250$ Hz (e.g., Fig. \ref{pic:SpecTime_R0R3}).
Comparing with the lower panels,
the spiral mode is seen to connect between the
standing shock front (the reddish high-entropy regions between the
outer blueish regions and inside) and the vicinity of the PNS ($\sim 50 -
60$ km in radius). In the next section, we are moving on to seek the
possible origins of the spiral waves in more detail.

\begin{figure}[htpb]
\begin{center}
\includegraphics[width=60mm,angle=-90]{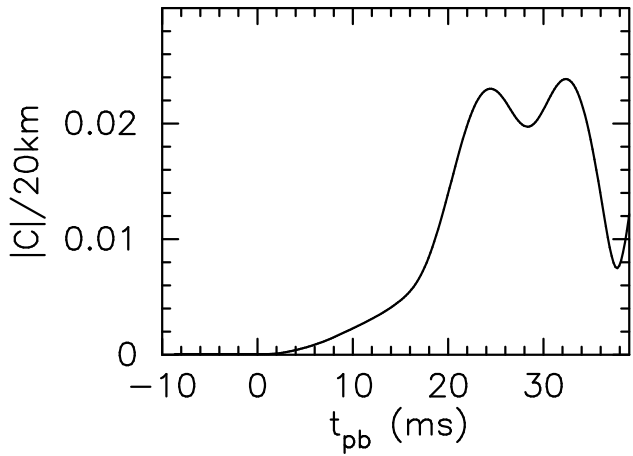}
\end{center}
\caption{\label{pic:mas_center} Postbounce evolution of 
the deviation of the mass center ($|C|=\sqrt{C_x^2+C_y^2+C_z^2}$) in the 
Cartesian coordinates from the origin for model R3. 
The relative displacement of the mass center is 
estimated at a fixed radius of 20km.}
\end{figure}

 Here we shortly mention that the appearance of the spiral wave is 
 unlikely to be seeded by non-conservation of momentum of the 3D 
hydrodynamic simulations. To check it, we plot in Fig \ref{pic:mas_center}
 that shows a postbounce evolution of the 
deviation of the mass center from the origin. The mass center 
($C_x,\ C_y,\ C_z$) in the Cartesian coordinates is defined as,
\begin{eqnarray}
C_i\equiv \frac{\int {\rho_\ast} x^i dx^3}{\int {\rho_\ast} dx^3}.
\end{eqnarray}
{\bf  Figure 16 shows the relative 
displacement of the mass center estimated at a fixed radius of 20km. 
 The mass center does not strictly stay in the very 
center, and the deviation maximally reaches
 $\sim 2 \%$ after $t_{\rm pb}\agt 10$ ms, when non-axisymmetric features
  are more clearly visible as shown in Figure 15. But 
the deviation never grows significantly in the simulation time 
($t_{\rm pb} \lesssim 30$ ms). Furthermore, since the absolute value of the displacement ($20~{\rm km} \times 0.02 \sim 0.4~{\rm km}$) is 
(albeit slightly) smaller than the finest grid size $\Delta {\rm x}=450$ m,
$C_i$  is located inside the innermost grid for every direction 
($i=x,y,z$). We thus consider that the momentum is conserved with a resolvable accuracy.}
 Note also that a vigorous growth of the one-armed spiral waves 
was observed also in the 3D post-Newtonian models including 
 a neutrino transport effect by \cite{Scheidegger10} around a similar 
postbounce timescale with ours (see their lower middle panel of Fig. 16).
Above facts indicate that non-axisymmetric instabilities observed in our simulations
do not come simply from a numerical artifact.
  
  \subsection{Possible Origins of One-armed Spiral Waves}
    In this section, we will discuss what kind of rotational instabilities
    took place, triggered the one-armed spiral waves and how they affect on the GW emissions.
    We consider two types of rotational instability may coexist which are
  (1) the low-$T/|W|$ instability \cite{Ott07_prl,Scheidegger10} which is mainly
  originated from the central PNS and (2) the spiral SASI
  \cite{Blondin07_nat,Iwakami08,Yamasaki08,Wongwathanarat10,Hanke13} which is originated from the stalled shock.
  We will discuss these instabilities more deeply in followings.
  
\begin{figure*}[hbpt]
\begin{center}
\includegraphics[width=65mm,angle=-90]{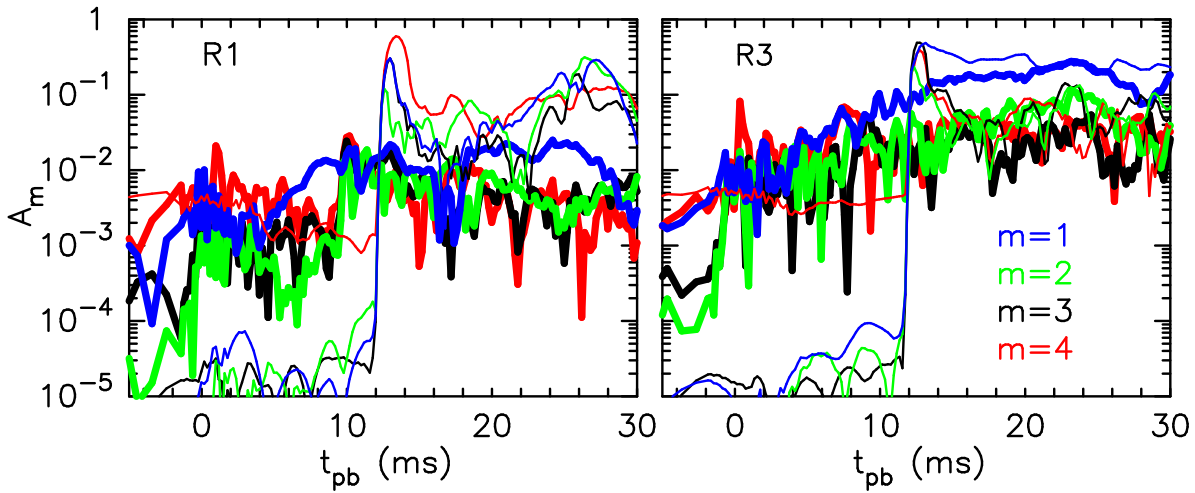}
\end{center}
\caption{\label{pic:AzimuthalMode} Time evolution of the normalized
 amplitudes of the density for
  different azimuthal modes ($m = 1,2,3,$ and 4) for models R1 ({\it
 left}) and R3 ({\it right}). Thick and thin lines are extracted at
  $\varpi=20$ and 130 km, respectively.}
\end{figure*}
In order to more clearly specify the non-axisymmetic structures,
we first monitor density profiles at a given radius $\varpi\equiv\sqrt{x^2+y^2}$ in the
equatorial plane in Fig. \ref{pic:AzimuthalMode}, which are evaluated by Eq. (\ref{eq:Am}).
For a mildly rotating model (R1, left panel of Figure \ref{pic:AzimuthalMode}),@the Cartesian $m=4$ background noise (red lines) shows relatively stronger signals than the other modes
inside the PNS ($\varpi = 20$ km, thick line)
 and at the shock ($\varpi = 130$ km, thin line) compared to rapidly rotating model R3.

We here shortly discuss whether these non-axisymmetric mode amplitudes 
within $T_{\rm pb}\alt10$ ms (e.g., $A_{\rm m}\sim10^{-2}$ for model R3)
are consistent with the resulting GW amplitudes ($\sim10$ cm) seen in 
Fig. \ref{pic:GWAmpMat} along the pole.
Based on an order-of-magnitude estimation, 
the GW amplitude (that we denote here as $A$) 
emitted from matter with mass quadrupole moment
$2M^2/R$ and with a measure of non-sphericity $\varepsilon$ can be roughly estimated as
\begin{eqnarray}
 \label{eq:Aestimate}
A\sim \varepsilon \frac{2M^2}{R}\sim \varepsilon \cdot 10^5\frac{M}{M_\odot} (\rm cm),
\end{eqnarray}
 where $M$ and $R$ represents the mass and size of the system, 
respectively.
 When we measure $\varepsilon$ from $A_{\rm m}$ and take
$M\sim 0.1M_\odot$ (typical value in the simulation time), 
the estimated GW amplitude becomes $\mathcal{O}(10)$ and 
$\mathcal{O}(10^2)$ cm for models R1 and R3, respectively.
 This is in good agreement with the obtained GW amplitudes,
 which suggests that percent levels of the non-axisymmetric mode 
amplitudes lead to sizable GW emission along the pole (e.g.,
Fig. \ref{pic:GWAmpMat}).
In addition, \cite{Ott07_prl} reported GW amplitudes $\sim10$ cm
with non-axisymmetric mode amplitude $A_{\rm m}\sim1$ \% (see, their Fig. 3) which is quantitatively consistent with our results.

 On top of the Cartesian
 noise, the linear growth of $m=1$ mode (inside the PNS for model
 R3, thick blue line) is clearly seen in the right panel, which
 gradually transits to the saturation phase as the prompt convection
 phase sets in ($t_{\rm pb}\agt10$ ms). Note that from
  the thin blue line (right panel, for $\varpi = 130$ km),
 the presence of the one-armed spiral waves behind the shock (compare
  Fig. \ref{pic:GW_loc}) is evident. 
  
  The dominant $m=1$ mode at the
  surface of PNS ($\varpi = 20$ km) is considered to be of a remarkable 
signature of the low-$T/|W|$ instability reported by, e.g., \cite{Ott07_prl,Scheidegger10}.
  Their studies showed that the frequency of the GWs 
associated with the instability typically peaks around $\sim$1 kHz.
 As seen from the top panel of Figure \ref{pic:dEdF_radial} (e.g., 
the spectrogram inside from $R \le 20$ km at $T_{\rm pb}\sim20$ ms),
there does exist some 
excess near 1 kHz before our simulations terminated (30 ms postbounce).
 But the excess is more clearly visible around $\sim$200 Hz, which 
 comes from outside the PNS. 
It takes typically several ten milliseconds after bounce before the low-$T/|W|$ instability fully develops enough to lead to a generation of higher order non-axisymmetric daughter modes $m\ge2$
(see Figure 3 in \cite{Ott07_prl}, and Figure 21 in \cite{Scheidegger10}).
After the $m=2$ deformation in the vicinity of the PNS, which is the most efficient GW emitter mode, a stronger GW emission with $\sim$1 kHz frequency
would be visible whose strength is also expected to overwhelm GWs emitted
from outer region $R \agt 60$ km.
In order to see much clearer features of 
 the low-$T/|W|$ instability, we should have continued
 our 3D models well beyond
 100 ms postbounce, i.e., till generation of higher order non-axisymmetric
 daughter modes, which is computationally expensive and is beyond 
the scope of this study.


Now we move on to the next discussion about one-armed spiral waves.
By carefully looking at the top panels (Fig.11), kinks are formed at the
triple points where the standing shock and the spiral wave meets.
Such type of morphology has been ubiquitously observed in previous simulations
 aiming to unravel the nature of the SASI
 \cite{Blondin07_nat,Blondin03,Scheck04,Ohnishi06,Foglizzo06,Yamasaki08,Iwakami09,
 Fernandez10}).
\begin{figure}
\begin{center}
\includegraphics[width=90mm,angle=-90]{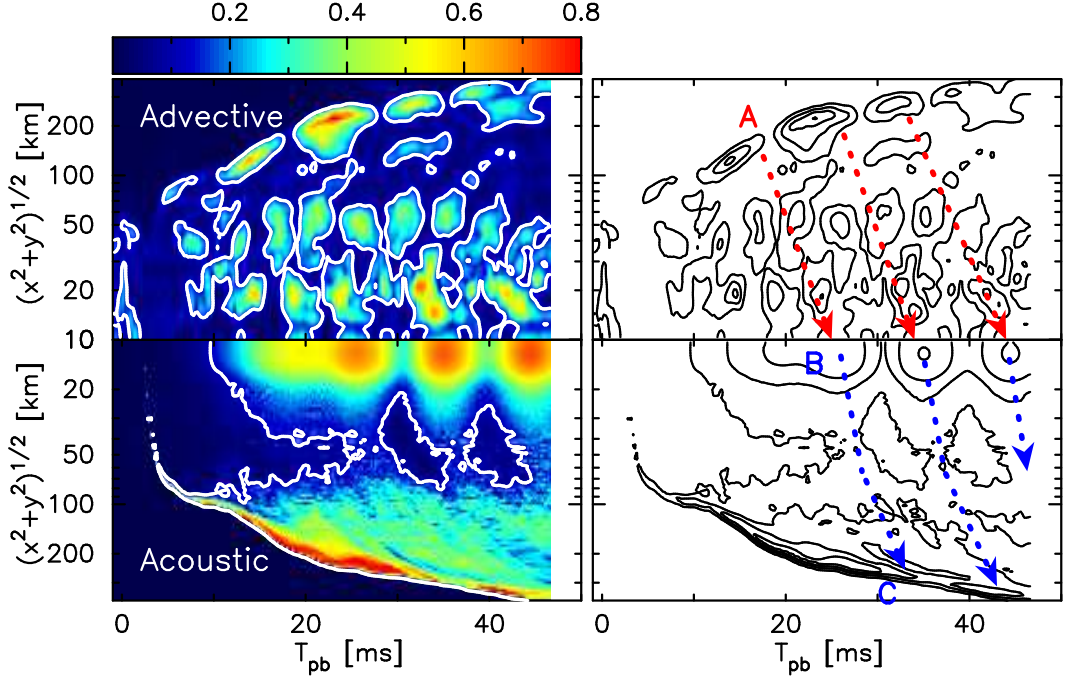}
\end{center}
\caption{({\it Left})Space-time diagrams of
 the one-armed spiral components of vorticity (top) and acoustic waves (bottom) (see text for more details).
({\it Right}) To guide the eyes, dotted arrows are inserted
 to illustrate the advective-acoustic cycle on top of the contour
 lines from the left panels.}
 \label{pic:Adv_Aco_M1}
\end{figure}
To check more carefully in our results whether the advective-acoustic
 cycle \cite{Foglizzo00,Scheck08} is running or not,
 we plot in Fig.\ref{pic:Adv_Aco_M1} the space-time diagram of vorticity
 (top panel, denoted by ``Advective'') and acoustic amplitudes (bottom
 panel, by ``Acoustic'' ) for model R3. Here the two quantities are
 evaluated as
\begin{eqnarray}
\tilde a(\varpi)=\frac{\int_0^{2\pi}X(\varpi,\phi,z=0)Y_{1,1}(\pi/2,\phi)d\phi}{\int_0^{2\pi}X(\varpi,\phi,z=0)Y_{0,0}(\pi/2,\phi)d\phi},
\end{eqnarray}
where $X= \varpi \nabla\cdot (v_\phi {\bf e}_\phi)$\footnote{Here ${\bf e}_\phi$ is a unit azimuthal vector} \cite{Scheck08}
and $X= p$ ($p$ the pressure) are for the vorticity and
 acoustic amplitudes, respectively, $\varpi=\sqrt{x^2+y^2}$, $\phi={\rm
 tan}^{-1}(y/x)$, and $Y_{l,m}(\theta,\phi)$ is the spherical harmonics.
Note that purely sloshing SASI modes were
studied in detail based on 2D simulations \cite{Scheck08}, but here we primarily
focus on the $(l,m)=(1,1)$ mode to extract the characteristic pattern of the
 one-armed spiral wave.

To guide the eyes, the right panel in Fig.\ref{pic:Adv_Aco_M1}
illustrates propagation of the advective (red lines) and acoustic
 waves (blue lines), respectively.
 Just behind the shock (see, point ``A'' in the
  top right panel), the generated vorticities go down to the PNS surface (e.g., close to the point ``B''). This is supported by
 the arranged direction (e.g., from top-left to down-right direction
  in the panel) of the bumpy islands, each of which
  is separated by black contour lines. The reddish regions at
  $R \sim 10-20$km after $T_{\rm pb} \sim 20$ ms (see the bottom left
  colormap) represent strong generation of the acoustic waves near
   the PNS. They propagate outward until they hit the
    shock. This is seen from the direction of the greenish stripes in
    the colormap (bottom left), which is symbolically drawn
    by blue lines in the bottom right panel. 
As indicated by red lines after point A (top right), the advective-acoustic cycle
 is in operation subsequently.
 Above features are in good agreement with
 \cite{Foglizzo00,Scheck08} (see references therein), although the SASI
 signatures are not as
 clearly discernible as in the previous 2D \cite{BMuller12b} or 3D simulations
 \cite{Iwakami09,Fernandez10} due to the short
 postbounce evolution in this study.

\begin{figure*}[hbtp]
\begin{center}
\includegraphics[width=90mm,angle=-90]{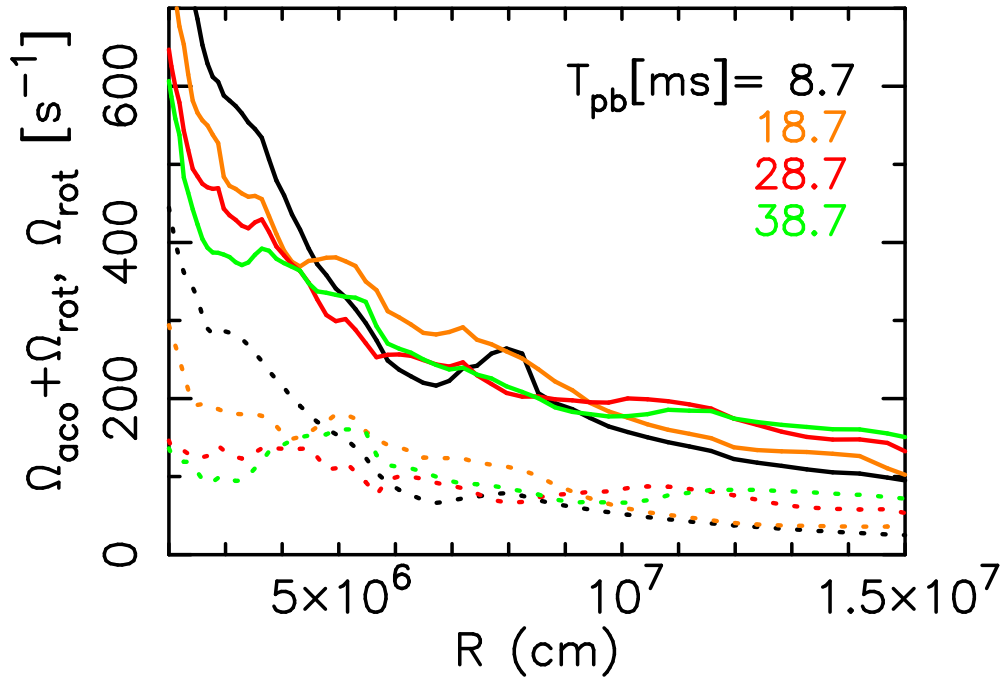}
\end{center}
\caption{\label{pic:RotationalAcoustic}  
  Profiles of $\Omega_{\rm aco}+\Omega_{\rm rot}$ (solid lines)
and $\Omega_{\rm rot}$ (dashed lines) along the equatorial direction ($x$
 axis) at different time slices (see text for more details).
}
\end{figure*}

In Fig.\ref{pic:Adv_Aco_M1}, stronger acoustic waves are generated
at $T_{\rm pb}\sim$27, 35, and 44 ms near the PNS approximately in the
time interval of $\sim$8 ms. This timescale is close to the
characteristic GW frequency of the spiral waves
($F_{\rm char}=200\sim 250$ Hz. e.g., in Fig.6). In order to better
address the origin of $F_{\rm char} \sim200$-250 Hz,
Fig.\ref{pic:RotationalAcoustic} shows angle-averaged frequencies
associated with the rotational velocity ($\Omega_{\rm rot}$, dotted lines)
and the acoustic wave ($\Omega_{\rm rot} + \Omega_{\rm aco}$, solid lines)
in the equatorial plane. Note that they are respectively defined as
\begin{eqnarray}
 \label{eq:Omega_rot}
\Omega_{\rm rot}&\equiv& 2\frac{V_\phi}{2\pi\sqrt{x^2+y^2}},\\
\label{eq:Omega_aco}
\Omega_{\rm aco}&\equiv& 2\frac{C_{\rm s}}{2\pi\sqrt{x^2+y^2}},
\end{eqnarray}
 and the sum ($\Omega_{\rm rot} + \Omega_{\rm aco}$)
 is a measure to estimate the propagation timescale of the acoustic
 wave on top of the rotating medium.
As seen, the acoustic frequency (solid lines) shows
$\Omega_{\rm aco}+\Omega_{\rm rot}\sim$200-250 Hz
in the range of $60$ km $\alt R\alt120$ km, which
is actually in good agreement with the narrow-band GW
emission seen in Figs. \ref{pic:dEdF_radial} and  \ref{pic:GW_loc}.
 In the regions above the PNS ($60$ km $\alt R\alt120$ km),
 it it also shown that $\Omega_{\rm rot}\sim 100$ Hz is as high as
 $\Omega_{\rm aco}$. The frequency difference ($\sim 100$ Hz) between the
 Doppler-shifted acoustic frequency
 ($\Omega_{\rm rot} + \Omega_{\rm aco}$) and the purely acoustic
 one ($\Omega_{\rm aco}$) can naturally explain the phase shift regarding
 the peak GW frequency
 between model R3 and the remaining
 models with smaller initial angular momentum. 

 Above results also show that the purely acoustic frequency
 $\Omega_{\rm aco}\sim100$ Hz is less sensitive to the initial
  angular momentum. In fact, the GW spectra in the literature
   \cite{Murphy09,Scheidegger10,BMuller13} generally peak at
   around $\sim100$ Hz during prompt convection. We thus speculate
   that significantly higher GW peaks\footnote{To
 make it possible, a correlation
 analysis with neutrino signals (e.g., \cite{Halzen09,Ott12a,Kotake12})
 should be indispensable to specify the epoch of core-bounce.},
 if observed in the spectrogram during prompt convection, might be a possible
 signature of rapid rotation, which is unobservable if not for the GW astronomy.
 
  \subsection{Detectability}
\begin{figure*}[htbp]
\begin{center}
\includegraphics[width=70mm,angle=-90]{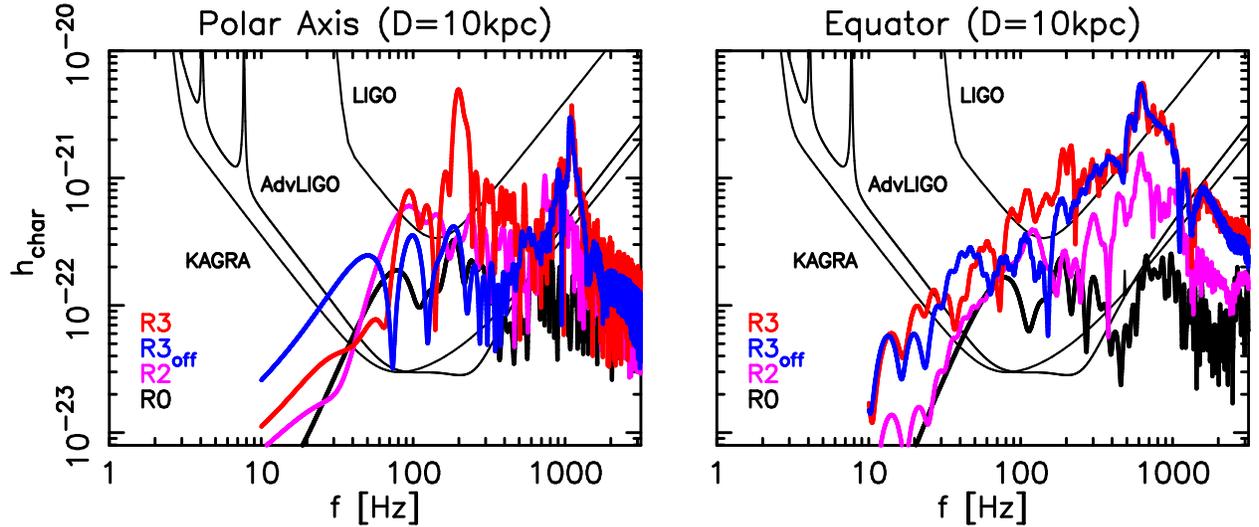}
\end{center}
\caption{\label{pic:GW_h_ch}  Characteristic GW amplitudes $h_{\rm char}$
 of our selected models (R3 (in red), R3$_{\rm off}$ (in blue), R2 (in magenda) and R0 (in black))
 compared with the strain sensitivity
 of initial LIGO\cite{Halzen09}, Advanced LIGO\cite{Harry10}, and
 KAGRA\cite{KurodaK10} at a source location of
 10 kpc. Left and right panels are for a spectator along polar and
 equatorial directions, respectively. Near the bottom left corner,
 the initial angular velocity of each model is provided for a reference.
  Note that due to the small contribution from the neutrino-originated
 GWs, only the matter contribution is plotted here.
}
\end{figure*}

  To discuss detectability, Fig.\ref{pic:GW_h_ch}
  shows the characteristic GW spectra $h_{\rm char}$ of our selected models
   with the design noise curves of initial LIGO
  \cite{Halzen09},  Advanced LIGO \cite{Harry10}, and KAGRA
  \cite{KurodaK10}), assuming a source distance of 10 kpc.

For our rapidly rotating models (e.g., R2 and R3), $h_{\rm char}$
  along the equatorial direction (right panel) is generally within the
 detection limits of the advanced detectors for an assumed distance of
  10 kpc. The corresponding
  signal-to-noise ratio (SNR) is approximately greater than
  $\sim$ 10 over a wide frequency range $100\alt F\alt 1000$ Hz.
In accord with previous 2D and 3D
  results\citep{Dimmelmeier02,Ott07_prl,Ott12a},
  faster initial rotation (but not too much) increase the chance for
  detection. The spectral peak appears at $F_{\rm peak} \sim620$ Hz for models R3
  and R2 (green and red lines in the right panel), which is associated
  with the maximum spike seen in the type I waveforms
   near bounce (e.g., Fig.\ref{pic:GWAmpMat} and the spectrogram in
   Fig.\ref{pic:SpecTime_R0R3}). Quantitatively this is in good
  agreement with Ott+07 \cite{Ott07_prl} who obtained
  $F_{\rm peak} \sim$600-700 Hz for their counterpart 3D model,
   in which the same Shen EOS was employed (but with much idealized
 microphysical treatment). 

Seen from the polar direction, the GW spectra have two distinct peaks
(e.g., green and red lines in the left panel of Figure \ref{pic:GW_h_ch}).
The first peak, appearing at $F_{\rm peak} \sim700$ Hz (model R2) and
$\sim1000$ Hz (model R3), is emitted during the neutronization phase
($t_{\rm pb}\alt 10$ ms).
As mentioned earlier, this component totally vanishes in
axisymmetry and it reflects the presence of precollapse density inhomogeneities.
 The second peaks are seen around $F_{\rm peak}\sim100$ Hz for
 non- to moderately-rotating models (red and black lines in the left
 panel). They are predominantly determined by the characteristic
  timescale of the acoustic waves traveling between the PNS and the stalled shock
  during prompt convection
 (see section \ref{nonaxis}). Due to rapid rotation,
 the acoustic frequency is shifted upward $F_{\rm peak} \sim200$ Hz
  in model R3 (green line in the left panel). As mentioned in the
  previous subsection, this is due to the
  Doppler effect (e.g., Eqns. (\ref{eq:Omega_rot}, \ref{eq:Omega_aco}))
  of the acoustic waves emitted on top of the spiral waves.
 The SNR of the first peak ($F_{\rm peak} = 1$kHz) for a Galactic source
 is at most SNR$\sim10$ even for model R3, whose sensitivity
 is limited by shot noise at high frequencies. On the other hand, the SNR of the
 second peak achieves as high as
 SNR$\sim100$ (for model R3), because it is close to
  the maximum detector sensitivity for the advanced interferometers.

\section{Summary and Discussions}
\label{sec:Summary and Discussions}
We have studied properties of GWs from
rotating core-collapse of a 15$M_\odot$ star by performing 3D-full-GR hydrodynamic simulations with 
an approximate neutrino transport. 
By parametrically changing the precollapse angular momentum, 
we paid particular attention to the effects of rotation on the 
GW signatures in the early postbounce evolution.
Regarding neutrino transport, we solved the 
energy-independent set of radiation energy and 
momentum based on the Thorne's momentum formalism. In addition to
the matter GW signals, we took into account GWs from anisotropic
neutrino emission. In addition to common GW signatures
obtained in previous 2D axisymmetric studies, our results showed
 several non-axisymmetric features in the waveforms
 which can be explored only by 3D simulations.

Among the common features are the type I waveforms emitted along
 the equatorial direction, which have comparable amplitudes
  with those in previous 3D GR studies \cite{Ott07_prl,Ott12a}.
The wave amplitude reaches a few $\times100$ cm for our most
rapidly rotating model R3. For a Galactic source,
this is well within the detection limits
(with the SNR $\gtrsim$ 10) over a wide frequency range
($100\alt F \alt 1000$ Hz) of KAGRA or Advanced LIGO.
 in our moderately to rapidly rotating models.
The peak GW frequency ($F_{\rm peak }\sim620$ Hz) in the GW
  spectra is also comparable to Ott+07 who employed the same Shen
EOS for hadronic matter.  During prompt convection,
the gravitational waveforms do not show any qualitative differences
 except for the most rapidly rotating model R3.
The wave amplitude is $|A|\sim10$ cm (non- and slowly rotating models)
and $|A|\sim20$ cm (moderately one)
which are consistent with previous GR simulations
\citep{Ott12a,BMuller13} and these signals reach the SNR$\sim8$,
independently of the observer direction.

We also studied neutrino luminosities, the average neutrino energies, and
the waveforms associated with anisotropic neutrino emission.
Rotation makes the neutrino luminosity and the average
 energies higher toward the polar direction, and conversely lower along
  the equatorial direction.
These features are consequences of rotational flattening of the
central core and neutrino spheres. The stronger
 neutrino emission toward the
 rotational axis leads to a quasi-monotonically increasing trend
  in the wave amplitude, as qualitatively similar to
  \cite{EMuller04} who employed more sophisticated neutrino transport.
 The impact of rapid rotation on the neutrino GWs
 was found only for the polarized wave mode $A_{+}$II in
 our most rapidly rotating model R3.

 Our findings clearly show that non-axisymmetric 
instabilities play an essential role in determining the GW signatures in 
the rotating postbounce evolution. 
By analyzing the GW spectrograms, the GW emission toward the rotational
axis has two distinct features, which appears at the higher
($F\sim700-1000$ Hz) and lower frequency domain ($F\sim100-200$ Hz), respectively.
The higher one near bounce, which appears clearly in our moderately and
rapidly rotating models, comes from the conventional 
rotating-bounce signal.
For a Galactic source, their signals achieve SNR$\sim5 - 10$ for the advanced
  detectors. The lower one seen in non- to moderately
rotating models is originated from spiral waves that develop
   under the control predominantly by the advective-acoustic cycle.
 In our most rapidly rotating model, the lower peak frequency
is shifted upward to $F\sim200$ Hz which is due to the Doppler effect 
  (e.g., Eqns. (\ref{eq:Omega_rot}, \ref{eq:Omega_aco}))
  of the acoustic waves emitted on top of the spiral waves.
 Regarding the detectability,
   the relevant GW frequencies are close to
  the maximum detector sensitivity for the second generation interferometers
  (e.g., KAGRA and Advanced LIGO), which is thus expected to be
   detectable (SNR$\sim 100$) for a Galactic source.

 Finally we would like to discuss some of the limitations of the present work.
For a more quantitative GW prediction, the simplifications in
 the employed neutrino reaction (and the gray transport scheme)
 should be improved, which we regard as the most
 urgent task (Kuroda, Takiwaki, and Kotake in preparation). We need to
 conduct a convergence check in which a numerical gridding is changed in
 a parametric manner, although it is too computational expensive to do
 so for our 3D-GR models at present. In addition, we need to run as
  many models as possible to study the dependencies of the precollapse
   density inhomogeneities, the progenitor dependence, the initial
   rotation rates and magnetic fields on the GW signals. Some good news
    is that we have access to the ``K-computer'' which is among
    the world-fastest Peta-scale
    supercomputers. By utilizing it, we hope to study
    these important themes one by one in the near future.

\acknowledgements
TK would like to thank continuous support from F.-K. Thielemann.
TT and KK are thankful to K. Sato for continuing encouragements.
KK acknowledges helpful discussions with Thierry Foglizzo, Christian
Ott, Rodrigo Fernandez, and H.T. Janka during the INT Program INT-12-2a
in University of Washington (2012). 
Numerical computations were carried out in part on 
XT4 and general common use computer system at the center for 
Computational Astrophysics, CfCA, 
the National Astronomical Observatory of Japan,
Oakleaf FX10 at Supercomputing Division in University of Tokyo, and on SR16000 at YITP in Kyoto University.
 This study was supported in part by the Grants-in-Aid for the Scientific Research 
from the Ministry of Education, Science and Culture of Japan (Nos. 20740150,
23540323, 23340069, and 24244036) and by HPCI Strategic Program of Japanese MEXT.

\bibliographystyle{h-physrev}
\bibliography{mybib}
\end{document}